\documentclass[12pt]{article}
\usepackage{graphics}
\usepackage{amssymb}
\usepackage{psfrag}

\newcommand{\rf}[1]{(\ref{#1})}
\newcommand{\beq}{\begin{equation}}
\newcommand{\eeq}{\end{equation}}
\newcommand{\bea}{\begin{eqnarray}}
\newcommand{\eea}{\end{eqnarray}}

\newcommand{\mi}{\!-\!}
\newcommand{\equ}{\!=\!}
\newcommand{\pl}{\!+\!}

\newcommand{\bose}{\mathbf e}

\newcommand{\be}{\begin{equation}}
\newcommand{\ee}{\end{equation}}
\newcommand{\ba}{\begin{eqnarray}}
\newcommand{\ea}{\end{eqnarray}}
\newcommand{\nn}{\nonumber}

\begin{document}

{\normalsize \hfill SPIN-05/15}\\
\vspace{-1.5cm}
{\normalsize \hfill ITP-UU-05/19}\\
${}$\\ 

\begin{center}
\vspace{48pt}
{ \Large \bf  Counting a black hole
\vspace{.4cm}\\
in Lorentzian product triangulations}

\vspace{40pt}

{\sl B. Dittrich}$\,^{a}$
and {\sl R. Loll}$\,^{b}$  

\vspace{24pt}
{\footnotesize

$^a$~Max-Planck-Institute for Gravitational Physics,\\
Am M\"uhlenberg 1, D-14476 Golm, Germany.\\
{email: dittrich@aei-potsdam.mpg.de}\\

\vspace{10pt}

$^b$~Institute for Theoretical Physics, Utrecht University, \\
Leuvenlaan 4, NL-3584 CE Utrecht, The Netherlands.\\ 
{email:  r.loll@phys.uu.nl}\\

\vspace{10pt}

}
\vspace{48pt}

\end{center}


\begin{center}
{\bf Abstract}
\end{center}
We take a step toward a nonperturbative gravitational path
integral for black-hole geometries by deriving an
expression for the expansion rate of null geodesic congruences
in the approach of causal dynamical triangulations. 
We propose to use the integrated expansion rate in building a
quantum horizon finder in the sum over spacetime geometries.
It takes the
form of a counting formula for various types of discrete building 
blocks which differ in how they focus and defocus light rays.
In the course of the derivation, we introduce the concept of a
Lorentzian dynamical triangulation of product type, whose
applicability goes beyond that of describing black-hole
configurations.

\vspace{12pt}
\noindent


\newpage

\subsection*{1. Introduction}\label{intro}

Very encouraging progress has been made recently in 
constructing {\it spacetime} dynamically from a nonperturbative gravitational 
path integral, by studying the continuum limit of causal dynamical triangulations 
\cite{al,ajl1,ajl3d,ajl4d}. 
The quantum geo\-metries generated in this way exhibit semiclassical
properties at sufficiently large scales: they are four-dimensional \cite{ajl-prl,univ}
and the large-scale dynamics of their spatial volume is
described by an effective cosmological minisuperspace action \cite{semi}. Their
short-distance behaviour is highly nonclassical, including a smooth dy\-na\-mical
reduction of the spectral dimension from four to two \cite{spec} 
and evidence of fractality \cite{univ}.

A question frequently asked of this and other approaches to quantum
gravity is what they have to say about the quantum dynamics of black holes and,
more specifically, whether the black-hole entropy ``comes out right". What is usually
meant by this is whether the theory can reproduce the Bekenstein-Hawking formula
$S\equ A/4$ which relates the entropy $S$ of a black hole to its area
$A$ in Planck units, and preferably at the same time provide a microscopic
explanation of the presence of black-hole entropy in terms of fundamental
excitations of geometry (see \cite{wald1,das,page} for recent reviews).
Although often portrayed as a touchstone for quantum gravity, it should be
kept in mind that the entropy formula and other thermodynamic relations
satisfied by black holes are semiclassical. 
Whether or not they have a fundamental role to play in a genuinely 
nonperturbative formulation of the theory remains to be seen.

This naturally raises the question whether the approach of
causal dynamical triangulations, which we believe is a strong candidate for
a nonperturbative theory of four-dimensional quantum gravity, can provide
new insights into the quantum properties of black holes. Apart from having
already reproduced certain classical aspects of general relativity from first
principles, the fact that the causal, Lorentzian structure of spacetime geometry 
plays a central role and that a well-defined Wick rotation is available 
at the regularized level seem to make this formulation particularly suited for 
addressing issues to do with black holes. However, putting this into practice turns 
out to be a challenging proposition. Not only are there technical obstacles to
be overcome, but one also has to decide   
what precise quantity should be calculated {\it if} one were given a well-defined
method for performing nonperturbative sums over geometries, such as causal
dynamical triangulations, a question that has hardly begun to be addressed. 

The main choices to be made when setting up the path integral are that of 
an ensemble of geometries to be summed over, and of boundary conditions
for the geometries. Since numerical evaluations
of the path integral must necessarily take place in a finite spacetime volume,
boundary conditions will have to be provided not only on some initial and final
spatial slice, but also on a (timelike) boundary at large radius, and potentially
also on a hypersurface at small radius to avoid any central 
singularity.\footnote{A detailed discussion of the inclusion of boundary
terms in causal dynamical triangulations can be found in \cite{kappel}.} 
We will for the moment set aside the question of
how the boundary conditions should be chosen,
and how many black holes one can expect to generate dynamically as a function 
of the boundary data. Instead we will focus on an issue that will be relevant
regardless of the geometric ensemble and boundary conditions chosen, namely,
which observable in the quantum theory can give us information about the presence 
or otherwise of a black-hole configuration. In a fully nonperturbative formulation,
this is a difficult task because of the absence of a classical background
structure. In the nonperturbative path integral, all possible geometries are
superposed, not just those which represent fluctuations around a given classical
background geometry. Semiclassical geometry emerges only in the
continuum limit and at sufficiently large scales \cite{semi,univ}. The problem is
then how and where in the dynamically generated quantum geometry one should look
for evidence of a black hole, say, an apparent horizon.

To simplify matters slightly, we will consider geometries with at most a single black
hole, and which moreover have (an approximate) spherical symmetry. To
translate these conditions to the setting of causal dynamical triangulations,
a useful concept is that of {\it triangulations of product type}. These are 
roughly speaking simplicial analogues of fibred spaces, with a triangulated
base space and triangulated fibres. The inspiration for such structures comes
in part from Lorentzian semi-random lattices \cite{fg}, which in turn were
motivated by causal triangulated models of quantum gravity in lower 
dimensions \cite{al,ajl3d}.\footnote{The presence of 
a M\"obius inversion formula for such
semi-random lattices, relating their partition function to that 
of a statistical model of geometric objects of one dimension less
(see also \cite{hop}), may help
to solve these models analytically. In the context of three-dimensional
quantum gravity in terms of causal dynamical triangulations, this possibility
is currently being explored \cite{dario}. A related quantum-cosmological
model which introduces more ``order" in three-dimensional random 
triangulations is described in \cite{hexagon}.} 
 
In the context of a nonperturbative description of four-dimensional black
holes, we propose to use triangulations of product type whose base
space is a simplicial version of the $r$-$t$-plane, with fibres representing
the two angular directions. For simplicity, it seems a good strategy 
not to include completely general geometries (triangulations) of this type in 
the path integral initially, but only a subset which satisfies certain 
homogeneity requirements along the angular directions, implementing
an approximate spherical symmetry at a coarse-grained scale. 
The fact that an exact continuous rotation symmetry cannot even in principle 
be realized by the simplicial structures we are working with is not
at all a drawback, but simply implies that the path integral will necessarily include
fluctuations which violate spherical symmetry, especially at short distances.
This is desirable from the point of view of the quantum theory, because it
is likely to make the model closer to the full theory than gravitational models
where an exact symmetry reduction is performed
{\it before} the quantization (see \cite{bojo} for an analogous reasoning in
canonical quantum gravity).   

In the present work, we will not attempt to define and evaluate a nonperturbative 
path integral for black holes, but pursue a more modest goal, namely,
to formulate a geometric observable for dynamical triangulations to help to
determine the ``presence" (in the sense of expectation values) of a black
hole in the quantum theory. The observable is a simplicial version
of the integrated expansion rate for light rays, whose vanishing 
in the classical continuum theory is an indicator
of the presence of an apparent horizon. We
derive an analogous quantity for causal dynamical triangulations which has
a particularly simple form, and which we hope can be used to construct an
efficient ``horizon finder" in Monte Carlo simulations of the quantum theory. 
It depends only on the numbers of
simplicial building blocks of various types and orientations occurring in
a given fibre of the product triangulation, but {\it not} on how these building
blocks are glued together locally. In this sense, the presence of an apparent
horizon can be established simply by counting.\footnote{To avoid any
misunderstandings, let us emphasize that this counting
has nothing to do with a counting of microstates to obtain an entropy for
a black hole.}

The rest of the paper is structured as follows. In Sec.\ 2 we introduce the general
concept of a Lorentzian triangulation of product type, before specializing to the
case of a two-dimensional base space. For product triangulations
in $2\pl 1$ and $2\pl 2$ dimensions, we classify all simplices occurring in
the fibres according to their orientation and derive some topological
relations among them. Sec.\ 3 is devoted to a discussion of extrinsic curvatures
and expansion rates. We start by recalling
the classical notion of a trapped surface and compute the light expansion
rates on a surface of codimension 2 from the extrinsic curvatures and other
geometric data. We then repeat the calculation for piecewise flat 
manifolds, by employing a careful limiting process at the ``kinks" of the
triangulation.\footnote{A related treatment in the context of classical Regge calculus
has appeared previously in \cite{brewin}.} This sets the stage for 
computing the integrated expansion rates for causal dynamical triangulations.
We derive the ``counting formulas", our main result, in both $2\pl 1$ and
$2\pl 2$ dimensions, and also give a qualitative interpretation 
in terms of the focussing and defocussing of light rays by particular
types of simplices. In Sec.\ 4, we construct an explicit example of a
triangulated black hole in the formalism of Lorentzian product
triangulations, and Sec.\ 5 contains a summary and outlook. The 
appendix deals with the definition of affine coordinates, which are used
in some of our derivations. --
Part of the work presented here is contained in the diploma thesis of one of
the authors \cite{bddiplom}.

\subsection*{2. Lorentzian triangulations of product type}

A triangulated manifold $\cal T$ of product type is a particular case of a
simplicial mani\-fold. Topologically speaking, it is 
a cartesian product ${\cal T}\equ B\times F$ of a
$b$-dimensional base space $B$ with an $f$-dimensional fibre
space $F$. A simplicial realization of this structure consists of a
simplicial base manifold $B$, where to each $k$-dimensional
simplex, $0\leq k\leq b$, ($k$-simplex for short) $\sigma$ of $B$ we associate a
so-called $\sigma${\it -tower}, which is a particular triangulation of $\sigma\times F$,
whose top-dimensional simplices are of dimension $k\pl f$. 
We will only consider triangulations ${\cal T}$
with a finite number of $d$-simplices.
For $k\equ b$, the dimension of the simplices in the $\sigma$-towers is
maximal, and coincides with the dimension $d\equ b\pl f$ of ${\cal T}$. 
The triangulations of these towers as well as the number of $d$-simplices
they contain will in general depend on the $b$-dimensional base simplex $\sigma$.  
In order that the $\sigma$-towers fit together in a simplicial manifold, neighbouring
towers have to satisfy matching conditions along their common $(d\mi 1)$-dimensional
boundaries, which themselves are $\sigma$-towers over the $(b\mi 1)$-dimensional
simplices of the base manifold $B$. The triangulations of the $d$-dimensional
towers induce triangulations on all lower-dimensional $\sigma$-towers, that is,
on the vertex towers, the edge towers, etc. By definition, each vertex of a
product triangulation lies in precisely one vertex tower.

The focus of our interest will be on Lorentzian triangulations of product type, and 
our main applications will have a two-dimensional base triangulation.
Note that the simplicial manifolds one considers in the approach of causal
dynamical triangulations can be thought of as a special class of 
triangulations of product
type, with base space $B$ the one-dimensional proper-time direction, and fibre
$F$ a spatial hypermanifold of constant time.
The simplicial realization of $B$ is simply a one-dimensional chain of time-like
edges, all of equal length. For $d$-dimensional
simplicial spacetimes, the spatial slices of constant integer time $t$ are the
$(d\mi 1)$-dimensional vertex towers over the (zero-dimensional) vertices of 
$B$, and the minimal spacetime ``sandwiches" between times $t$ and $t\pl 1$
are the edge towers over $B$. Each vertex of the spacetime 
triangulation lies in the vertex tower of exactly one vertex in $B$ and thus
inherits an integer time coordinate. 

\begin{figure}[ht]
\vspace{.5cm}
\centerline{\scalebox{0.6}{\rotatebox{0}{\includegraphics{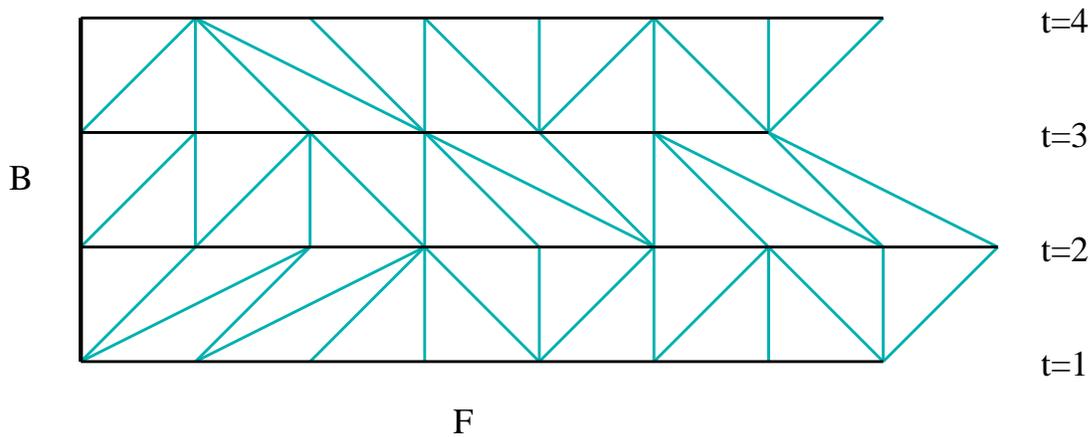}}}}
\caption[phased]{{\small
Example of a two-dimensional Lorentzian triangulation ${\cal T}$ whose base
space $B$ is the one-dimensional triangulation consisting of three
timelike edges. Note that in our graphical representation lengths are
{\it not} represented isometrically, in fact, this is impossible because of the 
intrinsic curvature carried by ${\cal T}$. There is only one type of triangular
building block which appears with either up- or down-orientation.
Consequently, all spacelike edges (horizontal) and all timelike edges
(interpolating between adjacent slices of constant integer time) have
identical length, although the latter fact is not rendered faithfully in 
the figure.
}}
\label{2dlor}
\end{figure}

The simplest non-trivial case of a causal triangulation is in dimension
two, an example of which is depicted in Fig.\ \ref{2dlor}. The vertex towers
are one-dimensional chains of spatial edges or links, and the edge towers
are linear sequences of two-dimensional triangles pointing up or down.
In a useful notation that generalizes to more complicated examples,
we denote an up-triangle by $[2,1]$ (``two vertices in the tower above the base
vertex at time $t$, one vertex in the tower above the base vertex at $t\pl 1$"),
and a down-triangle by $[1,2]$ (the converse). Each strip is uniquely characterized
by a sequence of such number pairs. Analogously, the $d$-simplices in a
minimal spacetime sandwich of a $d$-dimensional causal triangulation ${\cal T}$ can
be characterized by a pair $[i_1,i_2]$, $i_k\in\{1,\dots,d\}$, where $i_1$
counts the number of vertices at time $t$ and $i_2$ those at time $t\pl 1$ \cite{ajl4d}.
Since a $d$-simplex has $d\pl 1$ vertices, there are $d$ different ways how
its vertices can be distributed over the two constant-time slices, leading
to configurations of type $[d,1]$, $[d\mi 1,2]$, ..., $[1,d]$.

The product triangulations we will consider are
Lorentzian, that is, they are assembled from flat $d$-dimensional 
Minkowskian simplicial building blocks and have a product structure
with respect to the time direction. We show in the appendix that in such
manifolds the notion of a constant-time slice can be extended naturally 
to non-integer $t$. The resulting $(d\mi 1)$-dimensional hypersurfaces 
are again piecewise flat, but the individual building blocks are not
necessarily simplices (for example, the intersection pattern obtained
by cutting through a three-dimensional
Lorentzian triangulation is a simplicial manifold
consisting of triangles and rectangles). These concepts can be
extended straightforwardly to the case where the base manifold $B$
is higher-dimensional. Because of the Lorentzian structure 
of the triangulation, one of
the directions in $B$ will be the time direction. For example, in 
the simplicial analogues 
of spherically symmetric models discussed below, dim($B)\equ 2$,
with one time $t$ and one radial direction $r$. Also in this case, there
will be a well-defined notion of an $f$-dimensional fibre $F$ over an arbitrary
non-integer base point $(t,r)$. This fibre takes the form of a piecewise
flat manifold, with generalized flat building blocks.   
 
\begin{figure}[ht]
\vspace{.5cm}
\centerline{\scalebox{0.5}{\rotatebox{0}{\includegraphics{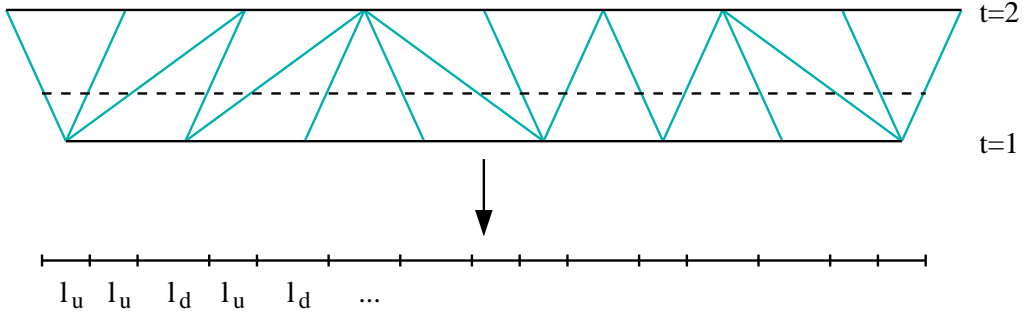}}}}
\vspace{.2cm}
\caption[phased]{{\small An edge tower $\sigma\times F$ of a two-dimensional 
Lorentzian triangulation can be represented by a one-dimensional graph
consisting of long and short edges 
which corresponds to a cut through the
tower at some time which does not coincide with $t\equ 1/2$. In the figure, 
the cut at $t< 1/2$ is indicated by the dashed line, the short edges
have length $l_u$ and the long ones $l_d$. Because of our planar
representation, the horizontal distances in the interior of the strip are
again not rendered faithfully.
}}
\label{strip1}
\end{figure}

\begin{figure}[ht]
\vspace{.5cm}
\centerline{\scalebox{0.5}{\rotatebox{0}{\includegraphics{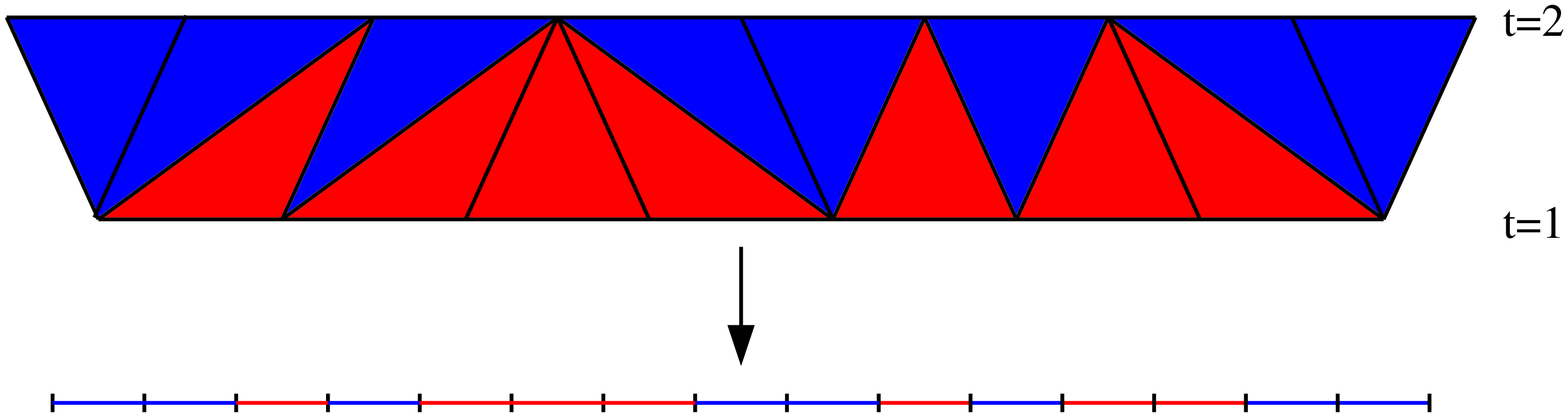}}}}
\vspace{.5cm}
\caption[phased]{{\small Alternatively, the same edge tower can be represented 
by a one-dimensional graph consisting of edges of two colours, which
can be thought of as being induced from colouring the up- and
down-triangles differently and considering a cut at midtime $t\equ 1.5$.
}}
\label{strip2}
\end{figure}

An important and useful observation is the fact that the $d$-dimensional
geometry of
a $\sigma$-tower, where $\sigma$ is a $b$-simplex in $B$, can be
deduced entirely from the geometry of the fibre $F$ over a single
interior point of $\sigma$. A simple example
is again given by a two-dimensional Lorentzian triangulation. The fibre
of constant time $s$ over some interior point $s$ of a strip $[t,t\pl 1]$, 
with $t$ an integer, is a sequence of straight edges. In order to be able
to distinguish the edges that come from cutting up- and down-triangles,
we must choose $s\not= t\pl 1/2$ (a similar exclusion of symmetric points
applies also in more complicated examples). This will result in edges of two
different lengths $l_u$ and $l_d$, as depicted in 
Fig.\ \ref{strip1}, and it is obvious that we can
reconstruct the entire strip (or $\sigma$-tower) from the knowledge of
the sequence of the two types of edges occurring in the fibre over $s$. 
An alternative way of keeping track of the two different types of edge
that can occur in the tower is to colour-code them (Fig.\ \ref{strip2}). Both
procedures generalize to more complicated product manifolds, with
each $\sigma$-tower represented uniquely by either a  (generalized)
$f$-dimensional triangulations or a special type of multicoloured graph 
(see the following sections for further examples, and the
appendix for geometric details),     
and the matching conditions for $\sigma$-towers taken into
account appropriately.

\subsubsection*{2.1 Lorentzian product triangulation with two-dimensional base}

In preparation for later sections, we will now describe the geometry of
Lorentzian triangulations ${\cal T}$ with a two-dimensional base space, for
$d\equ 2\pl 1$ and $d\equ 2\pl 2$. In both cases, the top-dimensional simplices of
$B$ are two-dimensional triangles, and ${\cal T}$ consists of triangle towers.
By assumption, the base manifold is itself of the form of a $(1\pl 1)$-dimensional
Lorentzian product triangulation $B\equ B'\times F'$, 
of the type usually considered in 
two-dimensional causal dynamical triangulations \cite{al} and depicted in
Fig.\ \ref{2dlor} above. Its base $B'$
is a one-dimensional triangulation consisting of timelike edges $\sigma'$, with
associated edge towers $\sigma'\times F'$ 
consisting of sequences of up- and down-triangles    
in the spatial direction. For our general discussion, it will not play a role whether
the topology of these strips is spatially open or closed.
Note that because of the physical interpretation of the triangulations as
causal spacetime geometries, there is no symmetry between the
time and spatial direction of $B$. Since the number of triangles in
an edge tower $\sigma'\times F'$ is in general a function of time, 
$B$ cannot be thought of as a product triangulation with the spatial
direction as its base. 

It follows that our simplicial manifolds have the form of ``staggered" 
product triangulations, which can be thought of as fibrations over $B$
or $B'$. Accordingly, their $d$-simplices can be characterized by how
they fit into either of these product structures. The information concerning
the fibration ${\cal T}\equ B'\times F'$ is simply the number pair $[i_1,i_2]$
which counts how many vertices of the simplex lie in either one of the
two adjacent slices of constant time $t$, as described above. From this
specification one can uniquely deduce the geometry of the $d$-simplex,
i.e. which of its edges are spacelike and which are timelike, and compute
all volumes and angles of the simplex and its subsimplices. 
Analogously,
the orientation of a $d$-simplex with respect to the fibration
${\cal T}\equ B\times F$ is specified by a triple of numbers $[j_1,j_2,j_3]$, 
which count how many of its $d\pl 1 $ vertices lie in each one of
the three vertex towers contained in $\sigma\times F$. 

\subsubsection*{2.2 The example of 2+1 dimensions}

\begin{figure}[ht]
\vspace{.5cm}
\centerline{\scalebox{0.6}{\rotatebox{0}{\includegraphics{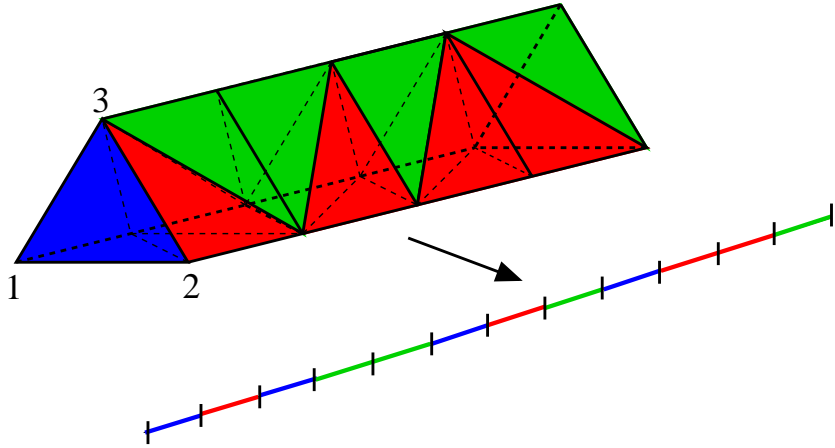}}}}
\vspace{.5cm}
\caption[phased]{{\small A triangle tower of a ($2\pl 1$)-dimensional
product triangulation can be represented by a one-dimensional
graph consisting of edges of three colours, corresponding to a
fibre over an interior point of the triangle. The three colours
correspond to the different orientations with which a tetrahedron can
appear in the prism. The figure depicts a sequence of tetrahedra
$[2,1,1]$, $[1,2,1]$, $[2,1,1]$, $[1,1,2]$, etc., where the labels of the
three vertex towers are as indicated.
}}
\label{prism}
\end{figure}

In order to train our geometric imagination, we start with the simpler
case $d\equ 3$  where the fibres $F$ are one-dimensional. 
From the point of view of the time fibration ${\cal T}\equ B'\times F'$,
the three-simplices or tetahedra come in three different
time orientations, $[3,1]$, $[2,2]$ and $[1,3]$. From the point
of view of the fibration ${\cal T}\equ B\times F$, there are three 
possibilities how a tetrahedron can appear in a tower above
a given base triangle $\sigma$, which in an obvious notation
are labelled by
$[2,1,1]$, $[1,2,1]$ and $[1,1,2]$. Consequently, one can
visualize the tower over $\sigma$ as a prism
with triangular base, which itself is of the form of a linear sequence
of tetrahedra of these three types. Assigning three different colours
to the differently oriented tetrahedra, we note that the geometry
of the triangulated prism can be encoded in a one-dimensional
graph made of edges of the three colours (Fig.\ \ref{prism}), namely, the graph 
corresponding to the fibre over any interior point of $\sigma$. 
This is a three-dimensional analogue of the situation depicted
in Fig.\ \ref{strip2}. 

In order to formulate the matching conditions between neighbouring
triangle towers in this representation, one needs to translate the
geometry of the edge tower common to the two triangle towers
into a one-dimensional graph. One simply considers the fibre
over an interior point of the edge in $B$ in question.
It is clear that this gives rise to
a two-coloured graph, because it represents a cut through a
two-dimensional triangulated strip. Denoting by
$\sigma_1$ and $\sigma_2$ the two triangles in
$B$ adjacent to the edge, such
a two-coloured graph is obtained by moving an interior point
in one of the $\sigma_i$'s toward the shared edge. In the process, one
type of coloured edge disappears from the fibre above the
point, because the corresponding type of tetrahedron shares
only an edge with the common edge tower, and not a triangle. 
The matching condition can therefore be formulated as a condition
on the linear graphs characterizing $\sigma_1\times F$ and 
$\sigma_2\times F$ with one colour each deleted.   

\begin{table}
\begin{center}
\renewcommand{\arraystretch}{1.4}
\begin{tabular}{ |c||c|c|c|}
\hline
simplex type $\backslash$ in graph with colours  &   $i_1i_2i_3$ &   $i_1i_2$ &  $i_1$  \\
\hline\hline
edge of colour $i_1$ & $[2_{i_1},1_{i_2},1_{i_3}]$  &  $[2_{i_1},1_{i_2},0_{i_3}]$ 
& $[2_{i_1},0_{i_2},0_{i_3}]$   \\
\hline
vertex & $[1_{i_1},1_{i_2},1_{i_3}]$  & $[1_{i_1},1_{i_2},0_{i_3}]$  & 
$[1_{i_1},0_{i_2},0_{i_3}]$  \\
\hline
\end{tabular}
\end{center}

\caption{\label{tab3d1} From this table, one reads off 
the unique building block $[j_{i_1},j_{i_2},j_{i_3}]$ in the triangle tower
of a $(2\pl 1)$-dimensional product triangulation 
corresponding to a simplex of a
one-dimensional coloured graph (left column). The building block
depends on the fibre type, as expressed by the number of colours
occurring in the coloured graph to which the simplex belongs (top row). 
The three $i_k$ are colour labels.}
\end{table}

We close this subsection with some further geometric observations,
which are straightforward in a three-dimensional context, but 
generalize to the more difficult four-dimensional case. First, one may also
associate a linear graph to a vertex tower, which can only ever
consist of edges of a single colour. Second, to this graph and
any other graph corresponding to a fibre over a point of $\sigma$
the Euler relation
\begin{equation}
N_0=N_1+\rho
\label{eulergr}
\end{equation}
must apply, relating the number $N_0$ of vertices in the graph
to the number $N_1$ of edges, irrespective of their colour. The
variable $\rho$ is either 0 or 1, depending on whether the graph
topology is $S^1$ or $[0,1]$. 
Table \ref{tab3d1} lists which building block in the triangle tower
corresponds to which element (vertex or coloured edge) of a
one-dimensional coloured graph, depending on whether the graph
comes from a fibre over an interior point of $\sigma$ 
(and thus has three colours $i_1$, $i_2$,
$i_3$), from a fibre over an edge (excluding the vertices) of $\sigma$
(with two colours, e.g. $i_1$, $i_2$),
or from a fibre over one of the three vertices (with a single colour) ,
e.g. $i$). For an interior point of $\sigma$,
relation (\ref{eulergr}) implies
\begin{equation}
N_{111}=N_{211}+N_{121}+N_{112}+\rho,
\label{tetrarel}
\end{equation}  
where the subscripts of the counting variables $N_{j_1j_2j_3}$ 
refer to the
characterization of tetrahedra in the triangle tower according
to the numbers $[j_1,j_2,j_3]$ of their vertices that lie in the vertex
towers 1, 2 and 3. Subsimplices of a given tetrahedron
inherit a triplet $[j_1,j_2,j_3]$ in an obvious way, with the three
$j_i$ adding up to the dimension of the subsimplex plus one.
This accounts for the appearance of the counting number of
triangles on the left-hand side of equation (\ref{tetrarel}).
Lastly, note that for the set of graphs associated with a given base
triangle, edges of a particular colour occur with the same
multiplicity in all graphs if the colour occurs in the graph at all.
This translates into the relation
\begin{equation}
N_{211}=N_{210}=N_{201}=N_{200}
\label{gleich}
\end{equation}
and permutations thereof. It follows that all numbers $N_{j_1j_2j_3}$ 
in one triangle tower are determined by $N_{211}$, $N_{121}$ and 
$N_{112}$. Moreover, these three numbers are independent, since
each type of tetrahedron can be inserted at any position of a
triangular prism. 

\subsubsection*{2.3 The example of 2+2 dimensions}

In comparison with the previous subsection, we will use the same
simplicial base space $B$, but increase
the dimension of the fibre $F$ from one to two. 
Let us call the spatial direction of
the base space the ``radial direction" (as will indeed be the case in
later sections).
The four possible types
of four-simplices from the point of view of the time fibration are
$[4,1]$, $[3,2]$, $[2,3]$ and $[1,4]$. 
These building blocks can still have different orientations 
within a given triangle tower.
Without loss of generality, let us assume that the three vertices of the 
base triangle $\sigma$ have coordinates $(t_1,r_1)$, $(t_1,r_2)$ and
$(t_2,r_3)$. For a $[4,1]$-building block, say, there are now
three possibilities how the four vertices in the vertex tower over the 
time $t_1$ (with respect to the fibration $B'\times F'$) can be distributed 
over the two vertex towers $(t_1,r_1)$ and $(t_1,r_2)$
(with respect to the fibration $B\times F$), labelled by
$[3,1,1]$, $[1,3,1]$ and $[2,2,1]$. Similarly, a $[3,2]$-building block
can occur with two orientations, $[2,1,2]$ and $[1,2,2]$, but there is only
a single way, $[1,1,3]$, to orient a $[2,3]$-building block. 

The question now arises of how the geometry of a triangle tower
can be captured by looking at the two-dimensional $B$-fibres over
(interior) points of its base triangle $\sigma$, which are cuts through 
${\cal T}$ of constant time and radius. The top-dimensional building blocks
of these surfaces are triangles (coming from $[3,1,1]$ and its
permutations) and rectangles (from $[2,2,1]$ and its permutations).
Like in previous examples, we can now colour-code subsets of 
edges in $F$ that will
always appear with a common length, independent of the base
point in $B$ of the fibre. Again they fall into three sets, which we will associate
with the colours red, green and blue. The triangular building blocks in
$F$ are monochrome, while the rectangles are all bi-coloured, with
opposite (and parallel) sides of identical colour. The fibre $F$ therefore takes the
form of a piecewise flat manifold with triangles and rectangles which
are glued together along edges of identical colour. 

\begin{table}
\begin{center}
\renewcommand{\arraystretch}{1.25}
\begin{tabular}{ |c|c|c|c|c|} 
\hline
simplex type & $B'\! \times\! F'$ type & fibre $F'$ & fibre $F$ & dual graph  \\
\hline\hline
$[3,1,1]$ & $[4,1]$ & tetrahedron & triangle rrr & 3-val.\ vertex r \\
\hline
$ [1,3,1]$  & $[4,1]$ & tetrahedron & triangle ggg & 3-val.\ vertex g   \\
\hline
$ [1,1,3]$ &  $[2,3]$ & prism & triangle bbb & 3-val.\ vertex b  \\
\hline
$ [2,2,1]$  & $ [4,1]$  &  tetrahedron & rectangle grgr & 4-val.\ vertex  gr\\
\hline
$[2,1,2]$ &$ [3,2]$ & prism & rectangle rbrb & 4-val.\ vertex rb \\
\hline
$ [1,2,2]$ & $ [3,2]$ & prism & rectangle gbgb & 4-val.\ vertex gb \\
\hline \hline
$[2,1,1]$r & $[3,1]$ &  triangle r& edge r & edge r \\
\hline
$[1,2,1]$g & $[3,1]$ &  triangle g &  edge g & edge g \\
\hline
$[1,1,2]$b & $[2,2]$ & rectangle b & edge b& edge b \\
\hline\hline
$[1,1,1]$  & $[2,1]$ & edge & vertex & polygon \\ \hline
\end{tabular}
\end{center}
\caption{ \label{tab4d2} Four-simplices in the triangle tower over a base triangle
with vertex coordinates $(t_1,r_1)$, $(t_1,r_2)$ and $(t_2,r_3)$, how they 
appear in the product triangulation $B'\times F'$ with time base, and in the fibres
$F'$ and $F$, together with a dual representation of the latter. The three colours
are indicated by $r$, $g$ and $b$.}
\end{table}

Alternatively, it is
sometimes convenient to use the graph dual to this (generalized) triangulation.
Instead of triangles we then have dual monochrome trivalent vertices,
and instead of the rectangles dual bi-coloured four-valent
vertices, consisting of pairs of mutually crossing edges of different colour.    
Since the subgraphs of a single colour close on themselves, the dual
graph has the form of a superposition of three trivalent monochrome
planar graphs. Similar to what happened in the $(2\pl 1)$-dimensional
example, if the base point of the fibre is an interior point of an edge of
$\sigma$, (dual) edges of one of the colours disappear, and if the base
point coincides with a vertex of $\sigma$, there are only (dual) edges of
a single colour left. 

Summarizing our findings, we can say that the geometries of triangle, edge
and vertex towers of a $(2\pl 2)$-dimensional product triangulation are
uniquely characterized by superpositions of three, two and a single
monochrome trivalent planar graph, respectively.\footnote{This picture is
reminiscent of three-dimensional causal dynamical triangulations,
where the three-dimensional `sandwich' geometry of a discrete time step  
$\Delta t\equ 1$ can be represented by a dual graph which is a superposition
of {\it two} monochrome trivalent planar graphs \cite{matrix}.} Table \ref{tab4d2}
summarizes the simplex types that can occur in a triangular tower, and
how they appear with respect to the various fibrations. 

\begin{table}
\begin{center}
\renewcommand{\arraystretch}{1.4}
\begin{tabular}{ |c||c|c|c|}
\hline
building block $\backslash$ in dual graph 
 &  $i_1i_2i_3$ & $i_1i_2$ & $i_1$  \\
\hline\hline
3-val.\ vertex $i_1$ & $[3_{i_1},1_{i_2},1_{i_3}]$ & $[3_{i_1},1_{i_2},0_{i_3}]$ 
& $[3_{i_1},0_{i_2},0_{i_3}]$   \\
\hline
4-val.\  vertex $i_1i_2$  & $[2_{i_1},2_{i_2},1_{i_3}]$  & $[2_{i_1},2_{i_2},0_{i_3}]$ &    \\
\hline
edge $i_1$   & $[2_{i_1},1_{i_2},1_{i_3}]$  &  $[2_{i_1},1_{i_2},0_{i_3}]$ & 
$[2_{i_1},0_{i_2},0_{i_3}]$   \\
\hline
polygon & $[1_{i_1},1_{i_2},1_{i_3}]$  & $[1_{i_1},1_{i_2},0_{i_3}]$  & 
$[1_{i_1},0_{i_2},0_{i_3}]$  \\
\hline
\end{tabular}
\end{center}
\caption{\label{tab4d1} From this table, one reads off 
the unique building block $[j_{i_1},j_{i_2},j_{i_3}]$ in the triangle tower
of a $(2\pl 2)$-dimensional product triangulation 
corresponding to a building block of a dual
planar coloured graph (left column). The building block
$[j_{i_1},j_{i_2},j_{i_3}]$
depends on the fibre type, as expressed by the number of colours
(three, two or one) 
occurring in the coloured graph to which the simplex belongs (top row). 
The three $i_k$ are colour labels. }
\end{table}

With this characterization in hand, we can now apply the Euler relations for
planar graphs,
\begin{eqnarray}
&&N_0^\star-N_1^\star+N_2^\star=\chi\nonumber \\
&&3 N_0^{(3)\star}+4 N_0^{(4)\star}=2N_1^\star,
\label{eulerdual}
\end{eqnarray}
relating the numbers $N_d^\star$ of building blocks of dimension $d$ 
contained in the dual graphs associated with a triangle tower. 
In (\ref{eulerdual}), $N_0^\star\equ N_0^{(3)\star}+ N_0^{(4)\star}$
is the sum of dual three- and four-valent vertices, $N_1^\star$
and $N_2^\star$ are the numbers of dual edges and faces, regardless
of their colour, and $\chi$ is the Euler number of the fibre. Because
there are different graphs associated with base points in the interior of
the triangle $\sigma$, and its edges and vertices, (\ref{eulerdual}) 
amounts to a total of 14 equations for the counting numbers $N_{j_1j_2j_3}$
for the various simplex types of a given triangle tower. They can be
used to express $N_1^\star$ and $N_2^\star$ as functions of 
$N_0^{(3)\star}$ and $N_0^{(4)\star}$, say. Applying (\ref{eulerdual})
to the graph representing the full triangle tower and making use of
the ``translation table", Table \ref{tab4d1}, we obtain
\begin{eqnarray}
&&N_{311}\pl N_{131}\pl N_{113}\pl N_{221}\pl N_{212}\pl N_{122}
\mi N_{211}\mi N_{121}\mi N_{112}\pl N_{111}=\chi, \nonumber \\
&&3 ( N_{311}\pl N_{131}\pl N_{113} )\pl 4 (N_{221}\pl N_{212}+N_{122}) =
2 (N_{211}\pl N_{121}\pl N_{112} ).
\label{eulerlong}
\end{eqnarray}
Furthermore, the numbers of three- and four-valent vertices of the same
colour(s) are equal for each of the graphs in which they appear, that is,
\begin{eqnarray}
&& N_{311}=N_{310}=N_{301}=N_{300},\nonumber\\
&& N_{221}=N_{220},
\label{idents}
\end{eqnarray} 
as well as permutations hereof. In conjunction with eqs.\ (\ref{eulerdual}),
they imply that each $N_{j_1j_2j_3}$ is determined by the six numbers
$N_{311}$, $N_{131}$, $N_{113}$, $N_{221}$, $N_{212}$ and $N_{122}$.
Moreover, within each triangle tower, these numbers are independent,
because one can alter them separately by applying local changes 
in the geometry.

\subsection*{3. Extrinsic curvatures and light expansion rates}

The main aim of this section is the derivation of an expression 
for the light expansion rate $H$ of a spacelike 
$(d\mi 2)$-dimensional surface $S$ of constant time and constant radius
in a simplicial product manifold of dimension four, in order to formulate
necessary criteria for the presence of black holes in the quantum theory.
We start by reminding the reader
of the geometric meaning of trapped surfaces. For reasons of
completeness, we then review the construction of the expansion rate in terms
of the extrinsic curvatures of $S$ and a suitably chosen spacelike hypersurface 
$\Sigma$ for the case of a smooth $d$-dimensional manifold $(M,g_{ab})$, following 
the treatment in \cite{york}. We then translate this to simplicial manifolds,
starting with a construction of the extrinsic curvatures. The expansion rate is
first computed for a three-dimensional triangulation, because the geometry
of the situation is closely analogous to that in four dimensions, but much
easier to visualize. We discuss the role of different types of simplicial
building blocks and their effect in terms of the focussing and defocussing of light
rays. As one would expect, the simplicial expressions for the expansion rates are 
subject to discretization ambiguities. We find that for specific choices of
how the piecewise flat surface $S$ traverses the top-dimensional building
blocks of the triangulation, the formulas for the expansion rates in both
three and four dimensions take on a particularly simple form which will be
useful in numerical simulations.

\subsubsection*{3.1 Trapped surfaces}

A central question in our attempt to set up a path integral for black-hole geometries
is how one may recognize the existence of a black hole region or a horizon in the 
quantum geometry. One difficulty is that the usual definition of a black hole region is  
both a classical concept and highly non-local
(for details, see \cite{wald} and references therein). 
To find the black hole region one has to know the entire spacetime manifold 
${\cal M}\equ (M,g_{ab})$ 
and to find the entire causal past $J^-(\mathfrak J^+)$ of future null infinity $\mathfrak J^+$.
The spacetime ${\cal M}$ is then said to contain a black hole region if $\cal M$ does {\it not} 
coincide with the causal past of future null infinity. What motivates this definition is
the fact that ``nothing can escape from a black hole'', and, in particular, nothing can escape 
to future null infinity. 
The event horizon is defined as the boundary of the black hole region, 
which in turn is defined as the complement of $J^-(\mathfrak J^+)$ in $\cal M$.

This definition is not suited for the formulation of a nonperturbative path integral in
terms of dynamical triangulations, because there one can only work with spacetimes of 
finite volume.  A more local criterion for a horizon or a black hole region is the existence of 
a so-called trapped surface. This also appears in other approaches to quantum geometry 
(see \cite{LQGentropy} and references therein), 
in classical general relativity in order to define ``(nearly) isolated'' 
horizons \cite{dyn}, as well as in ``horizon finders" in numerical relativity
(see, for example, \cite{horizon}).

\begin{figure}[ht]
\vspace{.8cm}
\centerline{\scalebox{0.33}{\rotatebox{0}{\includegraphics{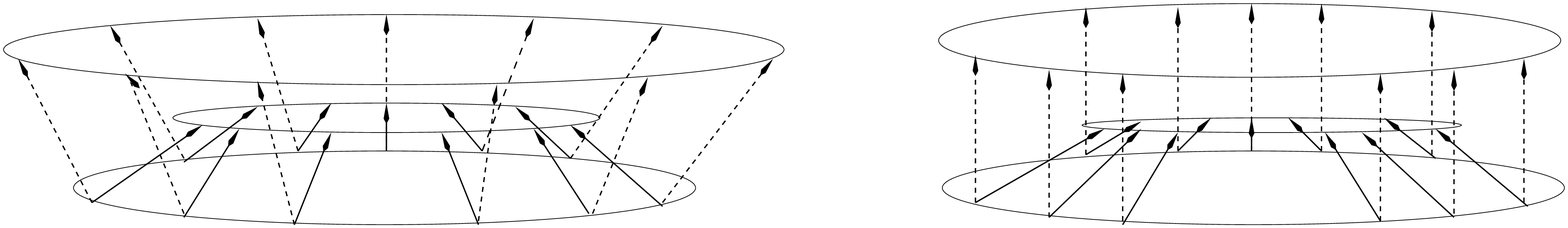}}}}
\vspace{.8cm}
\caption[phased]{{\small Inward- and outward-pointing light rays (solid and dashed arrows)
outside a black hole region (left) and at an apparent horizon (right). Here we depict the 
situation in a $(2\pl 1)$-dimensional spacetime, where the surface $S$ (lower circle) 
is one-dimensional. 
}}
\label{apphorizon}
\end{figure}
 
A trapped surface $S$ is a compact spacelike surface of codimension two, with the following 
property. Consider future-directed null geodesics (i.e. light rays), which start off orthogonally 
from $S$. This can happen in two directions, called ``inward-pointing'' and 
``outward-pointing'', see Fig.\ \ref{apphorizon}.
(The distinction is natural since $S$ is assumed to be closed, for instance, 
a sphere.) Light rays which point outward, say, can either diverge from or converge toward 
each other locally. This effect is measured by the outward-pointing expansion rate $H_+$, which 
can be either positive (for divergence), negative (for convergence) or vanishing. 
For a sphere in flat Minkowski space the expansion rate $H_+$ is everywhere positive and the 
analogous expansion 
rate $H_-$ for inward-pointing light rays negative. However, for a sphere inside a 
black hole, even the outward-pointing light rays are bent inwards to such an extent 
that the expansion rate $H_+$ becomes negative too.

Consequently, one defines a {\it trapped surface} as a compact spacelike surface of 
codimension $2$, whose expansion rates $H_+$ and $H_-$ are both negative. 
If ``negative'' is replaced by ``non-positive'' one speaks of a {\it marginally trapped surface}. 
The outermost marginally trapped surface is called the apparent horizon 
(assuming certain technical conditions, which are unimportant here). Its geometry is
depicted in Fig.\ \ref{apphorizon}.
One can show that its expansion rate $H_+$ vanishes everywhere.

For a Schwarzschild solution the apparent horizon coincides with the event horizon. 
Moreover, a marginally trapped surface 
is always contained in a black hole region, as defined above (see \cite{wald} and references 
therein for a precise formulation of the theorem). 
This holds also for spacetimes which do not fulfil the Einstein equations, but instead 
the condition $R_{ab}k^a k^b\geq 0$ for the Ricci tensor and all null vectors $k^a$. 
The latter is satisfied if the Einstein equations hold and if the matter satisfies 
the strong or weak energy condition.

These classical considerations motivate the replacement of the global characterization
of black holes in terms of event horizons by the more local criterion that a trapped surface exist. 
The condition $R_{ab}k^ak^b\geq 0$ will not in general be satisfied 
by the spacetime geometries contributing to the nonperturbative path integral, because
the individual path-integral histories are arbitrarily far away from classical solutions.  
However, one would expect to recover such a property at sufficiently large scales
in an appropriate continuum limit of the theory. We will in the following concentrate
on the condition $H_+=0$ as an indicator for the presence of black holes.

\subsubsection*{3.2 Continuum treatment}

The light expansion rate $H$ describes the expansion of a family of lightlike geodesics starting 
orthogonally from the $(d\mi 2)$-dimensional surface $S$. It is defined by 
\be \label{ler1}
H=\nabla_a p^a,
\ee
where $\nabla$ is the Levi-Civita connection associated to the spacetime metric $g_{ab}$ and $p^a$ is the tangent vector field to a future-pointing null geodesic congruence starting off orthogonally from the surface $S$. In other words, $p^a$ must satisfy
\ba \label{ler2}
& p_a p^a=0,\hspace{1cm} & m_{ab}p^a_{|S}=0,
\ea
and
\be\label{ler3}
p^a\nabla_a p^b=0,
\ee
where $m_{ab}$ is the metric induced on $S$ by the spacetime metric $g_{ab}$. 
Embedding $S$ into a spatial (constant-time) hypersurface $\Sigma$ with
future-pointing unit normal vector field $n^a$ and calling 
$\pm q^a$ the (out- and inward-pointing) unit normals to $S$ tangential to $\Sigma$,
the $(d\mi 2)$-dimensional metric is given by
\be\label{ler3a}
m_{ab}=g_{ab}+n_a n_b-q_aq_b.
\ee
For later use, we also introduce the notation
\ba  \label{ler6}
h_{ab}=g_{ab}+n_an_b \equiv m_{ab}+q_a q_b
\ea
for the induced metric on $\Sigma$.
Note that $m_{ab}$ is independent of the choice of $\Sigma$ as long as the latter contains $S$.
The general solution to eqs.\ (\ref{ler2}) on $S$ can then be written as
\ba\label{ler4}
{}^{\pm}p^a_{|S}=c_\pm(n^a \pm q^a),
\ea
with coefficients $c_\pm$ which are positive functions on $S$. 
As long as the functions $c_\pm$ are kept arbitrary, the general solution (\ref{ler4}) 
does not depend on $\Sigma$ in the sense that if one starts with another 
hypersurface $\Sigma'$ and corresponding normal vectors ${n'}^a$ and ${q'}^a$, 
one can always find functions $c'_\pm$ such that 
\be \label{ler5}
{}^{\pm}p^a_{\;|S}=c_\pm(n^a \pm q^a) =c'_\pm(n'^a \pm q'^a)={}^{\pm}{p'}^a_{\;|S}  .
\ee
Using the decomposition (\ref{ler3a}),
we can compute the expansion rates $H_{\pm}$ for the in- and outward pointing light rays, 
\ba   
\label{ler7}
H_\pm  &\!\!\! =\!\!\! & g^{ab} \nabla_a {}^\pm p_b \nn \\
       &\!\!\! =\!\!\! & m^{ab}\nabla_a {}^\pm p_b 
           +q^a q^b \nabla_a {}^\pm p_b 
           -n^a n^b \nabla_a {}^\pm p_b  \nn \\
       &\!\!\! =\!\!\! & m^{ab}\nabla_a {}^\pm p_b  
             -\left((c_\pm)^{-1}\,{}^\pm p^b \mp q^b\right) 
            (c_\pm)^{-1}\; \left[{}^\pm p^a \nabla_a {}^\pm p_b\right] 
           \pm q^a (c_\pm)^{-1}\; \left[ {}^\pm p^b \nabla_a {}^\pm p_b\right]
\nn \\
     &\!\!\! =\!\!\! & m^{ab}\nabla_a {}^\pm p_b   .
\ea
Going from the second to the third line, we have used eq.\ (\ref{ler4}).
The terms in square brackets vanish by virtue of relations (\ref{ler2}) and (\ref{ler3}). 
Noting that the covariant derivative is projected onto the surface $S$
in the last line of (\ref{ler7}), we can simplify the expansion rate further by using
again the expression (\ref{ler4}) for the vector fields ${}^\pm p^a$ on $S$,
\ba\label{ler8}
H_\pm  &\!\!\! =\!\!\! &  m^{ab}\nabla_a \, c_\pm (n_b \pm q_b) \nn \\
       &\!\!\! =\!\!\! &  c_\pm m^{ab}\nabla_a n_b  \pm c_\pm  m^{ab} \nabla_a q_b 
     +\left[ m^{ab}(n_b \pm q_b)\right] \nabla_a c_\pm  \nn \\
       &\!\!\! =\!\!\! & c_\pm \left(-  m^{ab} K_{ab} \mp k \right). 
\ea
The term in square brackets vanishes by definition of $p_b$, and in the third line 
we have defined the extrinsic curvature
\be\label{ler9}
K_{ab}=-h_a^c\nabla_c n_b
\ee 
of $\Sigma$ in $M$, and the extrinsic curvature 
\be\label{ler10}
k_{ab}=-m_a^c h_b^d \nabla_a q_b      
\ee 
of $S$ in $\Sigma$.

The last line of (\ref{ler8}) shows that the signs of the expansion rates $H_\pm$ on $S$ 
(which determine whether or not $S$ is a trapped surface) are independent of the 
prefactors $c_\pm$ introduced in eq.\ (\ref{ler4}) which are always positive.
Likewise, the condition $H_+\equ 0$ for an apparent horizon does not depend on 
the choice of the constant-time surface $\Sigma$ and therefore on $c_+$.
Our reason for keeping track of the prefactors $c_\pm$ explicitly is the fact
that in the quantum theory it may be convenient to monitor {\it integrated} 
expansion rates rather than local ones. If such an integration were performed
over a surface $S$ at a given time and radius and if the underlying geometry
were exactly spherically symmetric, the expansion rate would be 
constant on $S$ and 
the $\Sigma$-dependence of the integral would simply amount 
to an overall factor $c_\pm$. 
However, in situations without exact spherical symmetry, either in a smooth
or a simplicial setting, it does matter a priori that the absolute magnitude of 
$H_\pm$ depends on $c_\pm$. If we define an 
``averagely (minimally) trapped surface" $S$ in a generic geometry and for
a specific hypermanifold $\Sigma$ by 
\be 
\mathfrak{H}_+(S):=\int_S  H_+ \sqrt{\det m}\,{\rm d}S =0,
\label{average}
\ee
where $\sqrt{\det m}\,{\rm d}S$ is the invariant volume element on $S$, 
the same condition will in general not hold for a different choice $\Sigma'$
of the hypermanifold containing $S$, as was pointed out in \cite{hayward}.
The average expansion rate $\mathfrak{H}_+'$ may become positive, say,
because the prefactor $c_+'$ which appears when one expresses $p^a=n^a+q^a$ 
in terms of the primed normals,
\be\label{ler12}
p^a=c'_+(n'^a+q'^a),
\ee
may be large in a region of positive $H_+$ (defined with respect to $\Sigma$)
and small in a region of negative $H_+$, whereas the contributions from the
two regions
cancelled each other with respect to the unprimed spatial slice $\Sigma$.

It follows that some caution has to be applied when using the vanishing of
averaged expansion rates as an indicator for the presence of black holes.
This applies in particular to the case of simplicial manifolds which by their
very nature can never be exactly spherically symmetric.
Nevertheless, there are important situations -- including the one
considered here -- where the vanishing of the integrated
expansion rate, (\ref{average}), {\it is} the relevant criterion for indicating the 
presence of an apparent horizon 
in the quantum theory. Firstly, we may use the preferred time foliation
shared by all Lorentzian triangulated geometries in a quantum
superposition to define the null geodesic congruence in terms of eq.\  (\ref{ler4}) 
with $c_\pm \equiv 1$ and thus arrive at unique values for $H_\pm$.
Second, as explained in the Introduction, we are interested in studying the
path integral for approximately spherically symmetric geometries, in
which case the variations across $S$ of the expansion rate $H_+$ will not 
be large. Moreover, the surfaces $\Sigma$ will be approximately spherically 
symmetric, and therefore the factors $c'_\pm$ connecting the expansion rates of two 
different (approximately spherically symmetric) surfaces $\Sigma$ and $\Sigma'$ 
will be approximately constant, so that the case described above will generically 
not occur. 

One can also formulate (local or integrated) criteria for the presence of 
trapped surfaces which depend quadratically on the expansion rates $H_\pm$.
For example, a potential advantage of using the product $H_+ H_-$ is
the fact that it is independent of the choice of the constant-time surface
$\Sigma$ \cite{hayward}. A generic problem with such expressions in a
simplicial approach is that they contain products of delta functions at the
same point, and any regularization procedure is subject to a
high degree of discretization ambiguity. Also, one generically cannot
avoid an explicit dependence of the formulas on the local geometry of
the triangulation, as opposed to mere counting of simplex types, which can
be achieved for the linear expansion rates (c.f. Secs.\ 3.4
and 3.5 below). For these reasons, we will not pursue this possibility presently.

\subsubsection*{3.3 The case of piecewise linear manifolds}

In this section we will develop an expression for the expansion rate for the
case of a piecewise flat manifold, and will assume that both the constant-time surface
$\Sigma$ and the surface $S$ are fibres with respect to the two product structures 
involved. The second of these assumption we will relax later on in Secs. 3.4 and 3.5,
for reasons explained there.
In order to apply formula (\ref{ler8}) to the case of a 
piecewise linear manifold, 
we have to define an expression for the extrinsic curvature of a (piecewise 
linear) hypersurface in 
such a manifold. We begin by considering the extrinsic curvature $K_{ab}$ of a 
$(d\mi 1)$-dimensional constant-time hypersurface $\Sigma_t$ 
(see the appendix for a definition) in a $d$-dimensional triangulation. 
Since the geometry of each $d$-simplex is flat and Minkowskian, a constant-time 
surface inside it is just a linear spacelike hypersurface (with boundaries) of this 
Minkowski space, and its extrinsic curvature vanishes.  
We will arrive at the same result (\ref{plm11}) as reference \cite{brewin}.
However, our prefactors will differ from the ones obtained there, because we
use different coordinates, which also appear as argument of the delta-function.

A key observation is that one can always imbed two adjacent $d$-dimensional 
simplices $\sigma_1^d$ and $\sigma_2^d$ isometrically into a common $d$-dimensional 
Min\-kowski space. It follows that the geometry across the $(d\mi 1)$-dimensional 
boundary simplex $\sigma^{d-1}_{1 \cap 2}$ separating the two adjacent simplices is flat.  
(This no longer holds for the $(d\mi 2)$-dimensional subsimplices, on which the 
intrinsic curvature is concentrated.) In the common coordinate system, light rays crossing the 
$(d\mi 1)$-dimensional timelike boundary simplex
appear as straight lines, but a constant-time surface will in general be seen to have
a ``kink" there. This implies that its extrinsic curvature is concentrated on the $(d\mi 2)$-dimensional intersections $\Sigma_t \cap \sigma^{d-1}_{1 \cap 2}$ of the constant-time surface $\Sigma_t$ with 
the $(d\mi 1)$-dimensional timelike subsimplices. The relevant geometry is therefore
that of two linear spacelike hyperplanes in Minkowski space meeting in a kink (a 
$(d\mi 2)$-dimensional linear submanifold) with a certain relative angle. The geometry is 
homogeneous in the directions along the kink. In the triangulation, we can ignore these
directions as long as we stay away from the $(d\mi 2)$-dimensional boundary simplices 
of the $(d\mi 1)$-dimensional timelike subsimplex. 
\begin{figure}[h]
\psfrag{e0}{\Large$\bose_0$}
\psfrag{e1}{\Large $\bose_1$}
\psfrag{s1}{\Large $\mathbf{s}_{(1)}$}
\psfrag{s2}{\Large $\mathbf{s}_{(2)}$}
\psfrag{roh1}{\Large $\rho_1$}
\psfrag{roh2}{\Large $\rho_2$}
\psfrag{l}{\Large $x^1$}
\psfrag{l=0}{\Large $x^1=0$}
\psfrag{t}{\Large $t$}
\psfrag{n1}{\Large $\mathbf{n}_{(1)}$}
\psfrag{n2}{\Large $\mathbf{n}_{(2)}$}
\centerline{\scalebox{0.6}{\rotatebox{0}{\includegraphics{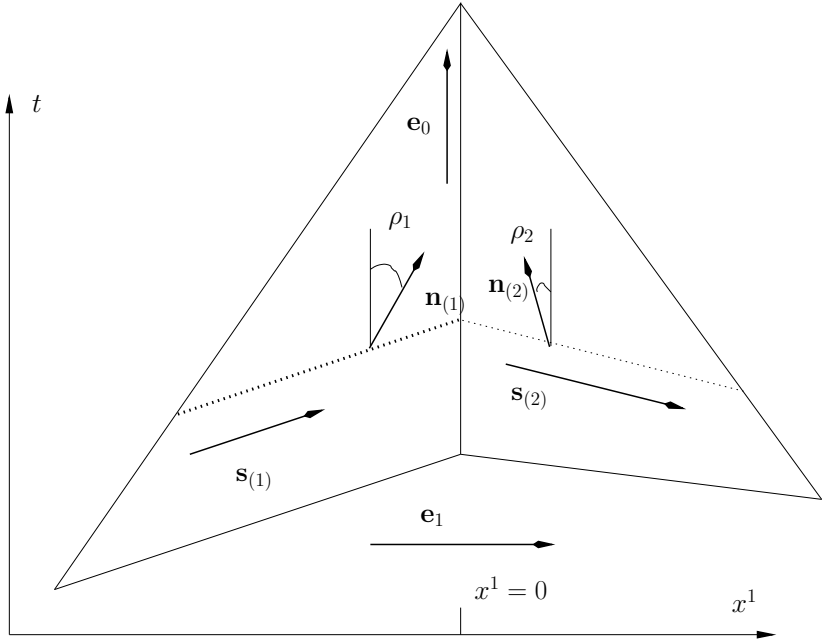}}}}
\caption{ \small The geometry of two neighbouring $d$-simplices, seen in the
two-dimensional plane perpendicular to the kink formed by the intersection of a
constant-time surface (dotted lines) and their common $(d\mi 1)$-boundary simplex.
The common Minkowski coordinate frame is spanned by the unit vectors 
$\bose_0$ and $\bose_1$, with corresponding coordinates $t$ and $x^1$. The
origin of the spatial coordinate is taken to coincide with the location of the boundary
simplex. Note that the normal vectors are perpendicular 
in a Lorentzian sense to the constant-time surface. 
\label{ExtrinsicCurvature}}
\end{figure}

To analyze the geometry of the two-dimensional plane spanned by the two vectors 
normal to the kink, it is convenient to use the basis vectors $(\bose_0)^a$ and 
$(\bose_1)^a$  orthogonal to $\Sigma_{t} \cap \sigma^{d-1}_{1 \cap 2}$ 
(see Fig.\ \ref{ExtrinsicCurvature}).
By convention, $(\bose_0)^a$ is future-directed and parallel to $\sigma^{d-1}_{1 \cap 2}$ 
and $(\bose_1)^a$ points from $\sigma^{d}_1$ to $\sigma^{d}_2$.
Furthermore, we introduce corresponding Minkowski space coordinates $(x^0,x^1)$, 
which measure the proper distances along the integral curves of $(\bose_0)^a$ and $(\bose_1)^a$
respectively. We define $x^1$ to be zero on $\sigma^{d-1}_{1 \cap 2}$.

For the computation of the extrinsic curvature tensor we need the unit normals 
$n^a_{(1)}$ and $n^a_{(2)}$ to the two pieces $\Sigma^{(1)}_t$ and $\Sigma^{(2)}_t$ 
of the constant-time surface. Because of their timelike nature, they can be written as 
\be \label{plm1}
n_{(i)}^a=\cosh\rho_i \, (\bose_0)^a + \sinh\rho_i \, (\bose_1)^a,
\ee
where $\rho_i, i=1,2$ is the hyperbolic angle between $n_{(i)}$ and $\bose_0$. Similarly,
the (spacelike) unit tangent vectors to $\Sigma^{(1)}_t$ and $\Sigma^{(2)}_t$ (in the span of $(\bose_0)^a$ and $(\bose_1)^a$) are 
\be\label{plm2}
s_{(i)}^a=\sinh\rho_i \, (\bose_0)^a + \cosh\rho_i \, (\bose_1)^a .
\ee
According to eq.\ \rf{ler9}, the extrinsic curvature is given by the projection onto $\Sigma_t$ 
of the covariant derivative of the normal vector field, which in our Minkowski coordinates 
reduces to the usual coordinate derivative. 
In order to deal with the discontinuity of the normal vector field of $\Sigma_t$ at the kink,
we regularize it with the help of a family of smooth functions $\delta_\varepsilon$ which 
converge to the delta function as $\varepsilon \rightarrow 0$. 
The angle $\rho$ between the time axis and the normal is then approximated by  
\be \label{plm3}
\rho(x^1)= \rho_1 + \Delta\!\rho \, \int_{-\varepsilon}^{x^1} \delta_{\varepsilon}(x'^1)\ {\rm d} x'^1 ,
\ee
where $\Delta\!\rho\equ\rho_2 \mi\rho_1$. Projecting $-\nabla_a n^b (x^1)$ 
onto the hypersurface $\Sigma_{t}$ with the projection operator $h_a^c:=\eta_a^c+n_a n^c$
(where $\eta_{ac}$ is the Minkowski metric), we obtain the extrinsic curvature
\ba 
\label{plm4}
K^{ab}(x^1)
&\!\!\!=\!\!\!& -(\eta^{ac} +n^a(x^1) n^c(x^1) )  \nabla_c  \big( \cosh\rho(x^1) \, (\bose_0)^b 
+ \sinh\rho(x^1) \, (\bose_1)^b \big) 
\nonumber \\
&\!\!\!=\!\!\!&-(\eta^{ac} +n^a(x^1) n^c(x^1) )  \delta_\varepsilon(x^1) \,\Delta\!\rho \, (\bose_1)_c \, s^b (x^1)   
  \nonumber \\
 &\!\!\!=\!\!\!& -  \delta_\varepsilon (x^1) \,\Delta \! \rho \, \cosh \rho (x^1)\, \, s^a (x^1)\, s^b(x^1)  ,
\ea
where we have used that $\nabla_a x^1=(\bose_1)_a$.

For the calculation of the extrinsic curvature $k^{ab}$ of the $(d\mi 2)$-dimensional surface 
$S$ in the surface $\Sigma_t$ we proceed completely analogously. 
The constant-time surface $\Sigma_t$ is a piecewise linear manifold, albeit a
generalization of a simplicial manifold with more general building blocks than simplices. 
Again, we can now embed any two adjacent $(d\mi 1)$-dimensional building blocks of the 
$\Sigma_t$-triangulation in a $(d\mi 1)$-dimensional flat euclidean space. 
We choose the surface $S$ as a linear $(d\mi 2)$-surface inside each building block, 
possibly with a kink when one crosses from one building block to the other. 

\begin{figure}[h]
\psfrag{e2}{\Large $\bose_2$}
\psfrag{s1}{\Large ${\mathbf s}_{(1)}$}
\psfrag{s2}{\Large ${\mathbf s}_{(2)}$}
\psfrag{u1}{\Large ${\mathbf u}_{(1)}$}
\psfrag{u2}{\Large ${\mathbf u}_{(2)}$}
\psfrag{psi1}{\Large $\psi_1$}
\psfrag{psi2}{\Large $\psi_2$}
\psfrag{l}{\Large $x^1$}
\psfrag{l=0}{\Large $x^1=0$}
\psfrag{r}{\Large $r$}
\psfrag{q1}{\Large ${\mathbf q}_{(1)}$}
\psfrag{q2}{\Large ${\mathbf q}_{(2)}$}
\centerline{\scalebox{0.6}{\rotatebox{0}{\includegraphics{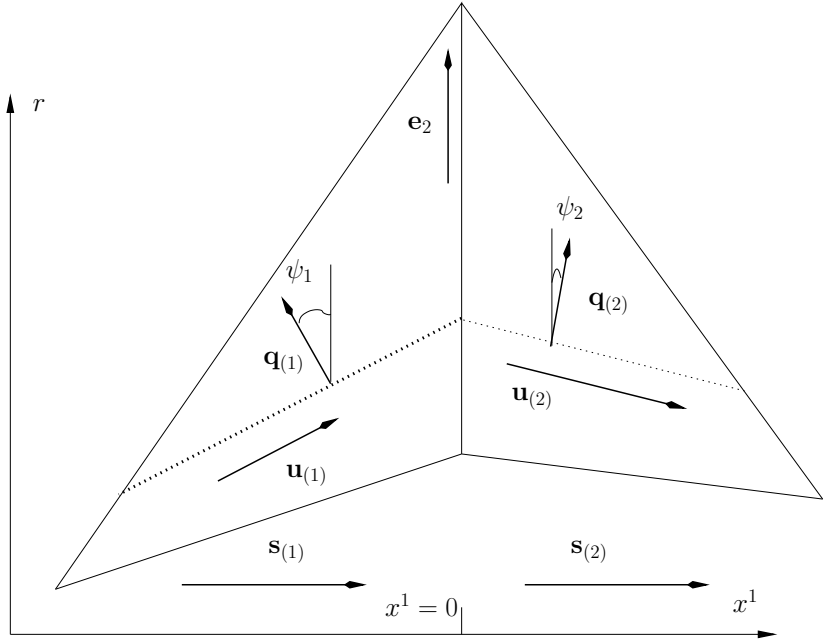}}}}
\caption{ \small The geometry inside a constant-time hypersurface $\Sigma_t$.
The figure illustrates various geometric quantities which appear in the calculation
of the extrinsic curvature of the surface $S$ (dotted line) in $\Sigma_t$. 
The surface $S$ has been chosen straight inside either of the two
neighbouring $d$-simplices, so that the only nontrivial contribution to the extrinsic
curvature comes from the kink of $S$ on the boundary between the two simplices.   
The vectors 
${\mathbf s}_{(1)}$ and ${\mathbf s}_{(2)}$ may differ in the direction of $\bose_0$ which 
is suppressed here.
\label{ExtrinsicCurvatureS}}
\end{figure}
To analyze the geometry of $S$, consider the same two $d$-simplices as before and
add a third basis vector $\bose_2$ which is defined to be orthogonal to $\bose_0$, $\bose_1$
and to the intersection $S \cap \sigma^{d-1}_{1\cap 2}$. 
For $d\equ 3$ the intersection is just a point, and for $d\equ 4$ the geometry is homogeneous 
along the intersection. In the latter case, we complement the basis with a fourth normalized 
vector $\bose_3$ orthogonal to the other three and thus tangential to $S \cap \sigma^{d-1}_{1\cap 2}$.

The (outward-pointing) unit normals $q^a_{(i)}$ to 
$S^i := S \cap \sigma^{d}_{i}$ in $\Sigma_t$
and those tangential vectors $u^a_{(i)}$ to $S^i$ which lie in the plane spanned by 
$\bose_2$ and $s_{(i)}$ may be written as
\begin{eqnarray} 
\label{plm5}
&& q^a_{(i)}=\; \;\;\;\cos \psi_i (\bose_2)^a + \sin \psi_i\ s^a_{(i)},  \nonumber \\
&& u^a_{(i)}= -\sin \psi_i (\bose_2)^a + \cos \psi_i\ s^a_{(i)} ,
\end{eqnarray}
where the angle $\psi_i$ is the angle between the normal $q_{(i)}$ and the
unit vector $\bose_2$.
Repeating our previous construction, we define an interpolating angle $\psi$ by 
\be
\label{plm6}
\psi(x^1)= \psi_1 + \Delta\!\psi \, \int_{-\varepsilon}^{x^1} \delta_{\varepsilon}(x'^1)\ {\rm d} x'^1 ,
\ee
with $\Delta\!\psi \equ \psi_2 \mi\psi_1$. 
To determine the extrinsic curvature of $S$, we begin by calculating
\begin{eqnarray}
\label{plm7}
&&{\rm D}_a q^b(x^1):= h_a{}\!^c(x^1) \, h^b{}\!_d(x^1) (\nabla_c q^d)(x^1) \nonumber \\
&&\hspace{1.8cm}= \delta_\varepsilon (x^1) \, \Delta \!\psi \cosh\rho(x^1) \, s_a(x^1)\,u^b(x^1)  ,
\end{eqnarray}
where ${\rm D}_a$ denotes the induced covariant derivative in the submanifold $\Sigma_t$.
This leads to the expression
\begin{eqnarray} 
\label{plm8}
&&k^{ab}(x^1)=-m^{ac}({\rm D}_c q^b)(x^1)=-(h^{ac}(x^1)-q^a(x^1)q^c(x^1))
({\rm D}_c q^b)(x^1)\nonumber \\ 
&&\hspace{1.4cm}= -\delta_\varepsilon (x^1) \, \Delta \!\psi \,\cosh\rho(x^1) \, \cos\psi(x^1)\, \, u^a(x^1)\,u^b(x^1)  .
\end{eqnarray}
for the extrinsic curvature of $S$ in $\Sigma_t$.
Given the extrinsic curvatures of $\Sigma_t$ and $S$, eqs.\ \rf{plm4} and \rf{plm8}, it is now 
straightforward to apply the continuum formula (\ref{ler8}) (with $c_+\equ 1$) and 
calculate the expansion rates $H_+$ of outward-pointing light rays as
\begin{eqnarray}
\label{plm9}
&&H_+(x^1)= -k - m^{ab}K_{ab} \nonumber \\
&&\hspace{1.4cm}= \delta_\varepsilon (x^1) \bigl(\Delta \!\psi\, \cosh \rho(x^1) \cos\psi (x^1) + 
\Delta\!\rho \, \cosh \rho (x^1)  \cos^2\psi (x^1) \bigr) .
\end{eqnarray}

For the computation of the integrated expansion rate, it is convenient to introduce the
variable $p$ measuring the proper length along the orbits of $u^a$. From
\be
\label{plm10}
u^a = -\sin \psi (\bose_2)^a + \cos \psi (\sinh\rho \, (\bose_0)^a + \cosh\rho \, (\bose_1)^a)
\ee
we have ${\rm d} x^1=\cosh \rho \cos \psi\, {\rm d} p$, and it follows that
\begin{eqnarray}
\label{plm11}
&&H_+(p) = \delta_\varepsilon (p) \bigl(\Delta \! \psi + \Delta\!\rho \cos\psi (p) \bigr) \\
&& \hspace{1cm}{\buildrel {\varepsilon \rightarrow 0_+}\over \rightarrow}  \;
 \delta(p) \bigl(\Delta \! \psi + \Delta\!\rho \cos\psi_m \bigr)
\end{eqnarray}
where $\psi_m:=\frac{1}{2}(\psi_1 +\psi_2)$ and we have assumed the functions $\delta_\varepsilon$ 
to be symmetric.
For the case of a four-dimensional triangulation, by introducing a second (proper-length)
coordinate $x^3$ along $(\bose_3)^a$ the invariant volume element 
$\sqrt{\det m}\,{\rm d}S$ of $S$ assumes the simple form ${\rm d} p\, {\rm d} x^3$. 
We can now easily integrate the expansion rate over a neighbourhood in $S$
enclosing the set $S \cap \sigma^{d-1}_{1 \cap 2}$ (that is, the support of the expansion rate in $S$),
resulting in
\be
\label{plm12}
\int_{{\rm nbh.}}  H_+ {\rm d}p\, {\rm d} x^3 =
{\rm Vol}(\sigma^{{\rm d}-1}_{1 \cap 2} \cap S) \bigl(\Delta \! \psi + \Delta\!\rho \cos\psi_m \bigr).
\ee 
(For a three-dimensional triangulation, $d\equ 3$, we set 
${\rm Vol}(\sigma^{{\rm d}-1}_{1 \cap 2} \cap S)\equ 1$). 
In the integral over the whole surface $S$, the contributions from the $(d\mi 3)$-dimensional 
building blocks simply add up to give the integral $\mathfrak{H}_+(S)$ 
of the expansion rate $H_+$ over the surface $S$, 
\be 
\label{plm13}
\mathfrak{H}_+(S)=\int_S  H_+ \sqrt{\det m}\,{\rm d}S=
\sum_{\sigma^{{\rm d}-1}\cap S} {\rm Vol}_{(\sigma^{{\rm d}-1}\cap S)} 
\bigl(\Delta \! \psi + \Delta\!\rho \cos\psi_m \bigr)_{(\sigma^{{\rm d}-1}\cap S)} .
\ee

When evaluating an expression like (\ref{plm13}) for a dynamical triangulation 
it is inconvenient that the angles $\rho_i$ and $\psi_i$ defined by eqs.\ (\ref{plm1}) and (\ref{plm5}) 
depend on the indexing of the $d$-simplices $\sigma_1$ and $\sigma_2$. 
This motivates the definition of new angles $\tilde{\rho}_i=(-1)^i \rho_i$ and $\tilde{\psi}_i=(-1)^i \psi_i$, 
in terms of which we have
\begin{eqnarray} 
&&\!\!\cosh\tilde\rho_i =- (\bose_0)^a\, n^b_{(i)} \eta_{ab}, \hspace{.4cm}
\sinh\tilde\rho_i= (\tilde\bose_1)^a_{(i)}\, n^b_{(i)} \eta_{ab}, \hspace{.4cm}
\Delta\! \rho=\tilde\rho_2+\tilde\rho_1\label{rhoangle} \\
&&\cos \tilde \psi_i = (\bose_2)^a \,q^b_{(i)}\, \eta_{ab},  \hspace{.9cm}
 \sin \tilde \psi_i =  \tilde s^a_{(i)}\,q^b_{(i)}\,\eta_{ab}, \hspace{.9cm}
\Delta\!\psi = \tilde\psi_2 + \tilde\psi_1 ,\label{psiangle}
\end{eqnarray}
where $(\tilde\bose_1)^a_{(i)}=(-1)^i(\bose_1)^a$ and $\tilde s^a_{(i)}=(-1)^i s^a_{(i)}$ 
are pointing toward the simplex $\sigma_i^d$ for both $i\equ 1$ and $i\equ 2$. 
Furthermore, $\cos\psi_m$ may be written as 
\be 
\label{plm15}
\cos \psi_m=\cos(\frac{1}{2}(\tilde\psi_1-\tilde\psi_2))=\cos(\frac{1}{2}(-\tilde\psi_1+\tilde\psi_2)) ,
\ee
so that all quantities $\Delta\! \rho$, $\Delta\!\psi$, $\cos\psi_m$  appearing in the integrated
expansion rate (\ref{plm13}) 
are independent of the index $i$ in $\sigma_i^d$.

\subsubsection*{3.4 The case of dynamical triangulations: 2+1 dimensions}

In order to get an idea of what is involved in applying the framework of the previous section
to the particular case of causal dynamical triangulations, we will first consider the simpler
case of 2+1 dimensions. The triangulations take the form of fibrations over
a two-dimensional base manifold, parametrized by a time and a radial coordinate.
Our aim will be to derive a formula for the (integrated) expansion
rate which is operationally simple to evaluate. 
In 2+1 dimensions the surface $S$ on which we want to define the expansion rate $H_+$ 
is one-dimensional and $H_+$ is concentrated on the zero-dimensional vertices 
$v=\sigma^{2}\cap S$ of the piecewise straight surface $S$. Following standard convention,
we set the volume of these vertices to $1$.
\begin{table}
\begin{center}
\renewcommand{\arraystretch}{1.6}
\begin{tabular}{ |c|c|c|}
\hline
simplex   & boundary simplex &  $\tilde{\rho}$   \\
\hline \hline
$[3,1]$ & $[2,1]$  &  $ -{\rm arsinh} \frac{1}{\sqrt{3+12\alpha}}$     \\
\hline
$[1,3]$ & $[1,2]$  & $  {\rm arsinh}\frac{1}{\sqrt{3+12\alpha}} $    \\
\hline
$[2,2]$ & $[2,1]$  & $  {\rm arsinh}\frac{1}{\sqrt{1+4\alpha}}  $    \\
\hline
$[2,2]$ & $[1,2]$  & $  -{\rm arsinh}\frac{1}{\sqrt{1+4\alpha}}  $     \\
\hline
\end{tabular}
\end{center} 
\caption{ \small  \label{Winkel1} The angles $\tilde{\rho}$ per simplex type
and per boundary simplex type contributing to the expansion rate 
in 2+1 dimensions. As usual in causal 
dynamical triangulations, we fix all spacelike edges
to have a squared length $l_s^2\equ a^2$ and all timelike edges to have
$l_t^2\equ -\alpha a^2$, where $a$ is the so-called lattice spacing that will
eventually be taken to zero. The number $\alpha$ parametrizes the (fixed)
ratio between the length units in time and space directions.}
\end{table}
Consider now a fibre $S$ in a $[1_{T,in},1_{T,out},1_{T+1}]$-triangle 
tower.\footnote{Here $T$ is an integer-valued time, and the 
lower-case $t$ will henceforth 
denote a time value between $T$ and $T+1$. The subscripts $in$ and $out$ refer to 
vertices with radial coordinates $r_{in}$ and $r_{out}$, with $r_{in}<r_{out}$.}
Applying formula (\ref{plm13}) for the integrated expansion rate gives
\be \label{211}
\mathfrak{H}_+(S)=- \mathfrak{k}(S)+\sum_{v \subset S}  \Delta\!\rho(v) \cos\psi_m(v) ,
\ee
where
\be \label{212}
\mathfrak{k}\,(S)=\frac{\pi}{3}\,(N_{211}-N_{121}) 
\ee
is the integrated extrinsic curvature of $S$ in $\Sigma_t$. The values of the
angles $\Delta\!\rho(v)=\tilde\rho_1(v)+\tilde\rho_2(v)$ and 
$\cos(\psi_m)(v)=\cos(\frac{1}{2}(\tilde\psi_1(v)-\tilde\psi_2(v)))$ contributing to 
eq.\ \rf{211} can be read off the Tables \ref{Winkel1} and \ref{Winkel2}. 
The term \rf{212} is just a counting term which only depends on the total number of simplices
of a certain type. This is not true for the remaining sum over vertices $v$ in eq.\ \rf{211}.
To evaluate it, we need to know not only the numbers of the various simplex types but also 
how they are glued together. In other words, there exist pairs of triangulations with identical
numbers $N_{xyz}$ which nevertheless have different (integrated) expansion rates. 
This dependence on the local gluing information is caused by the factor $\cos\psi_m$ in 
(\ref{211}), and can be traced back to our particular choice of the surface $S$. 
In the previous subsection, we assumed $S$ to be straight inside each 
(constant-$t$ hypersurface of a) three-simplex, 
as illustrated in Fig.\ \ref{ExtrinsicCurvatureS}. 
\begin{table}
\begin{center}
\renewcommand{\arraystretch}{1.6}
\begin{tabular}{ |c|c|}
\hline
$ [i_{ in,T} ,j_{ out,T},k_{T\pm 1}]$  & $\tilde{\psi}$  \\
\hline \hline
$[2,1,1]$ &    $ -\frac{\pi}{6}$     \\
\hline
$[1,2,1]$ &   $ \frac{\pi}{6}$    \\
\hline
$[1,1,2]$ & $ 0$  \\
\hline
\end{tabular}
\end{center}
\caption{ \small  \label{Winkel2} The angles $\tilde{\psi}$ per simplex type
in a triangle tower of type $[1_{T,in},1_{T,out},1_{T\pm 1}]$ contributing to the
expansion rate in 2+1 dimensions.}
\end{table}
This leads to a coincidence of angles
$\tilde\rho$ and $\tilde\psi$ at the intersection point $S \cap \sigma^{d-1}_{1\cap 2}$ 
between two three-simplices $\sigma_1^3$ and $\sigma_2^3$.
We will adopt a different choice for the surfaces $S$ which will simplify the
functional form of the integrated expansion rate. We expect that this will speed up an eventual 
numerical implementation and also simplify any analytic considerations of the
evaluation of the expectation value of the expansion rate in the path integral.
The precise choice of $S$ is a typical discretization ambiguity, which should
not make any difference when we perform the continuum limit.\footnote{Obviously such
an assertion must eventually be verified.} In the absence of an argument 
for which discretized choice of a ``surface of constant radius"
is more natural, we will adopt the simplest prescription, both here and in the
four-dimensional case discussed in the next subsection.

\begin{figure} 
\psfrag{[2,1,1]}{\large $[2,1,1]$}
\psfrag{r}{\large $r$}
\psfrag{[1,2,1]}{\large $[1,2,1]$}
\psfrag{[1,1,2]}{\large $[1,1,2]$}
\centerline{\scalebox{0.8}{\rotatebox{0}{\includegraphics{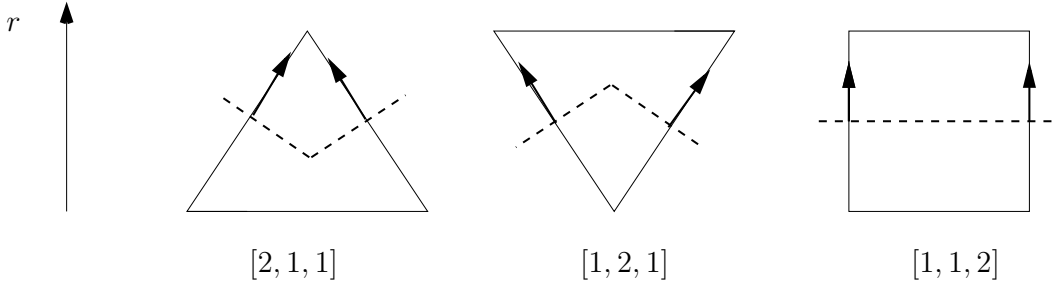}}}}
\caption{ \small  \label{faelleexprate} The ``building blocks" which constitute
our surfaces $\hat S$ of constant radius (dashed lines) inside the hypersurfaces 
$\Sigma_t$ in 2+1 dimensions. }
\end{figure}
Our alternative hypersurfaces $\hat S$ lie still within $\Sigma_t$, and are piecewise
straight, but we arrange their straight segments to be dual to the edges of the original
triangulation in $\Sigma_t$, as depicted in Fig.\ \ref{faelleexprate}, so that the
``kinks" of $\hat S$ no longer coincide with the points $S \cap \sigma^{d-1}_{1\cap 2}$.
Since the angles $\Delta \psi$ are still the same as for the original surface $S$,
the first term in the expansion rate \rf{211} is unaffected, 
$\mathfrak{k}(\hat S)=\mathfrak{k}(S)$. However, since $\cos\psi_m$ is now 
always equal to $1$, we obtain for the integrated expansion rate of $\hat S$
\begin{eqnarray} 
\label{213}
&&\mathfrak{H}_+(\hat{S})= -\frac{\pi}{3}\,(N_{211}-N_{121}) -  
\bigl\{ 2\,{\rm arsinh}\frac{1}{\sqrt{3+12\alpha}}\,\,(N_{211}+ N_{121})\,\bigr\} +\nonumber \\
 &&\hspace{2cm} 2\,{\rm arsinh}\frac{1}{\sqrt{1+4\alpha}}\,   N_{112}  
\end{eqnarray}
for a triangle tower of type $[1_{T,in},1_{T,out},1_{T+1}]$. 
For a triangle tower of type $[1_T,1_{T+1,in},1_{T+1,out}]$ the first term, i.e. the extrinsic 
curvature of $\hat S$, is unchanged (apart from relabelling), whereas the terms 
coming from the extrinsic curvature of $\Sigma_t$ change sign, leading to
\begin{eqnarray}
\label{214}
&&\mathfrak{H}_+(\hat{S})= -\frac{\pi}{3}\,(N_{121}-N_{112}) + 
\bigl\{ 2\, {\rm arsinh}\frac{1}{\sqrt{3+12\alpha}}\,\,(N_{121}+ N_{112})\,\bigl\} -\nonumber \\
&&\hspace{2cm} 2\,{\rm arsinh}\frac{1}{\sqrt{1+4\alpha}}\,   N_{211}  .
\end{eqnarray}
One thing to note about eqs.\ \rf{213} and \rf{214} is that the two terms in curly
brackets have merely boundary character in the sense that they cancel
when one considers the expansion rate for more than one time step. 
For example, suppose we added the integrated expansion rates associated with two 
base triangles of type $[1_{T-1},2_T]$  and $[2_T,1_{T+1}]$ with a common spacelike edge.
Then the contributions from the two terms in curly brackets cancel each other,
because the numbers of $[1_{T-1},3_T]$-tetrahedra in the first triangle tower equals the number of 
$[3_T,1_{T+1}]$-tetrahedra in the second. 

\begin{figure} 
\psfrag{t}{\LARGE $t$}
\psfrag{r}{\LARGE $r$}
\psfrag{angle}{\LARGE $\psi$}
\centerline{\scalebox{0.32}{\rotatebox{0}{\includegraphics{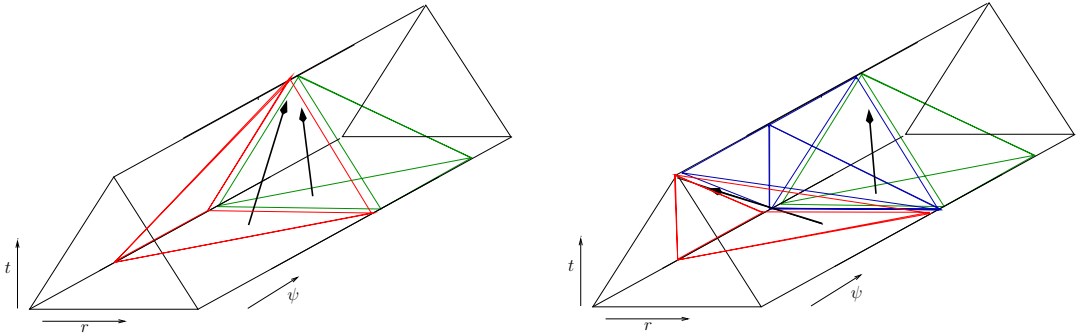}}}}
\caption{ \small  \label{3dexpraten2} The effect of different three-simplices in a
$[1_{T,in},1_{T,out},1_{T+1}]$-triangle tower on future-directed light rays.
We only show the angular behaviour, which is relevant for calculating the 
expansion rate. Tetrahedra of type [3,1] focus light rays in the angular
direction (left figure), whereas [2,2]-tetrahedra defocus them (right figure).
}
\end{figure}
A second observation about the counting formulas \rf{213} and \rf{214} is that they have 
a direct geometric interpretation in terms of the focussing of light rays passing through
the three-dimensional building blocks of the product triangulation. 
Let us call the fibre direction angular 
and the other spatial direction radial, and
consider first a $[1_{T,in},1_{T,out},1_{T+1}]$-triangle tower. 
It contains both [3,1]-tetrahedra (which can be split
into the subtypes [2,1,1] and [1,2,1]) and [2,2]-tetrahedra (of subtype [1,1,2]).
The first term in (\ref{213}), which depends on the difference $(N_{211}\mi N_{121})$,
has a purely spatial origin and is determined by the triangulation of the spacelike edge tower. 
The second term (in curly brackets) in \rf{213} gives a negative contribution, since a
[3,1]-tetrahedron focusses light rays in the angular direction (see Fig.\ \ref{3dexpraten2},
left). It also focusses light rays in the radial direction. At any rate, since these effects
cancel out for adjacent time slices as we have just explained, let us concentrate on
the remaining term in \rf{213}. As illustrated in Fig.\ \ref{3dexpraten2}, the
$[1,1,2]$-tetrahedra defocus light in the angular direction (thus accounting for
their positive contribution to the integrated expansion rate on $\hat S$), and
focus it in radial direction.
In a $[1_T,1_{T+1,in},1_{T+1,out}]$-triangle tower the focussing effects are exactly reversed: 
the $[1,3]$-simplices defocus light rays (or timelike normal vectors) whereas the 
$[2,1,1]$-tetrahedra focus in angular and defocus in radial direction. 
We conclude that apart from the extrinsic curvature term $\mathfrak{k}$ of the 
(one-dimensional) surface $\hat S$ the expansion rate is determined by 
simply counting the numbers of focussing and defocussing [2,2]-tetrahedra,
associated with the $[1_{T-1},2_T]$ and $[2_T,1_{T+1}]$-base triangles respectively,
and taking their difference.

\subsubsection*{3.5 The case of dynamical triangulations: 3+1 $\equ$ 2+2 dimensions}

We will now generalize the treatment of the previous subsection to the case of
a four-dimensional Lorentzian (i.e. causal) dynamical triangulation which is a product of
a two-dimensional triangulated ``$r$-$t$-plane" and a two-dimensional fibre,
representing the angular directions. We will follow the same strategy as in the
three-dimensional case to obtain a simple functional form for the expansion
rate $\mathfrak{H}_+(S)$ which does not depend on detailed local gluing data. 

For a $(2\pl 2)$-dimensional triangulation, $S$ is two-dimensional and $\mathfrak{H}_+(S)$
has distributional support on the one-dimensional edges $e=\sigma^3 \cap S$ in $S$. 
As before, if one defines the surface $S$ to be a two-dimensional fibre in a triangle tower, 
it turns out that the expansion rate does not just depend on the numbers of the various 
simplex types in the triangle tower, because of the $\cos\psi_m$-terms in 
eq.\ (\ref{plm13}). We will therefore choose an alternative surface $\hat S$,
still inside the constant-time hypersurface $\Sigma_t$, but which 
intersects the two-dimensional building blocks contained in $\Sigma_t$ orthogonally
such that all $\cos\psi_m$-factors will be equal to one. 

Note that in the $(2\pl 1)$-dimensional case the surface $\hat S$ was related to the 
dual triangulation of $\Sigma_t$ -- the vertices of $\hat S$ were positioned at the 
barycentres of the triangles and rectangles which are the piecewise flat
building blocks of $\Sigma_t$. (The vertices inside the rectangles did not appear
explicitly, because they do not contribute to the extrinsic curvature of $\hat S$.)
Also in 2+2 dimensions we will construct the surface $\hat S$ as part of the dual 
triangulation of the constant-time surface $\Sigma_t$. 
Recall from Sec.\ 2.3 that $\Sigma_t$ consists of
equilateral tetrahedra and triangular prisms with 
two equilateral triangular and three rectangular sides. To construct the dual 
triangulation, one places vertices at the barycentres of these tetrahedra and prisms,
which one then joins by ``dual" edges if they belong to neighbouring building blocks.
These dual edges will orthogonally cross the boundary faces (triangles or rectangles
shared by adjacent tetrahedra and/or prisms) exactly at their barycentres. 
The dual flat two-surfaces spanned between pairs of 
dual edges will therefore also intersect the two-dimensional faces 
of the $\Sigma_t$-triangulation orthogonally. Our surfaces $\hat S$ will by
definition be built from those dual two-surfaces which extend in the angular directions. 

\begin{figure}[h]
\psfrag{in}{\large{in}}
\psfrag{out}{\large{out}}
\centerline{\scalebox{0.6}{\rotatebox{0}{\includegraphics{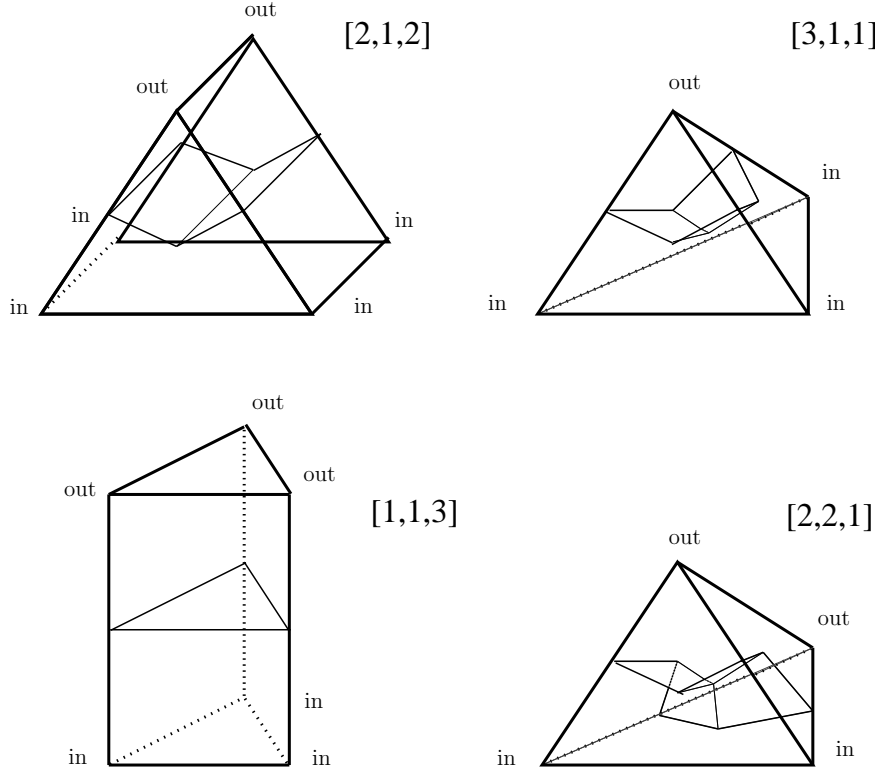}}}}
\caption{ \small \label{2plus2slicedb} The elementary contributions of various
three-dimensional building blocks in $\Sigma_t$ to the two-dimensional
surface $\hat S$ in a $[1_{T,in},1_{T,out},1_{T+1}]$-triangle tower. 
The cases not shown, associated with four-simplices of type [1,2,2] and [1,3,1],
can be obtained by exchanging the role of the ``in'' and ``out'' vertices of
the cases [2,1,2] and [3,1,1] respectively.}
\end{figure}
The extrinsic curvature $k$ of a surface $\hat S$ with respect to $\Sigma_t$ is now 
concentrated on the dual edges. For some dual edges $k$ vanishes, for example, for
the dual edges in the $[2_T,3_{T+1}]$-building blocks in a $[2_T,1_{T+1}]$-triangle tower. 
We will now discuss each case in turn and compute its associated angle contribution 
$\Delta\psi$.
Fig.\ \ref{2plus2slicedb} shows the three-dimensional flat building blocks
that can appear in a constant-time hypersurface $\Sigma_t$, each with its
corresponding ``elementary" surface contributing to $\hat S$. 
There are essentially four different cases:
\begin{itemize}
\item[(i)] The intersection of a four-simplex of type [2,1,2] or [1,2,2] with $\Sigma_t$
is a prism. The surface $\hat S$ intersects both triangular faces and two of the
rectangular faces. The dual edges in $\hat S$ which cross the rectangular faces 
do not carry any extrinsic curvature but the two edges which cross the triangular 
faces do. The angular difference between the two normals involved is 
$\Delta\psi=\pm \pi/3$. 
\item[(ii)] The intersection of a four-simplex of type [3,1,1] or [1,3,1] with $\Sigma_t$
is a tetrahedron. The surface $\hat S$ intersects three out of the four boundary
triangles. Three dual edges connect the barycentre of the tetrahedron with the centres 
of these three triangles. The angular difference for each of the three dual edges is 
again $\Delta\psi=\pm \pi/3$.   
\item[(iii)] The intersection of a four-simplex of type [1,1,3] with $\Sigma_t$
is a prism. The surface $\hat S$ intersects the three boundary rectangles. 
Since the intersection surface is a single flat triangle parallel to the
triangular boundaries of the prism, there are no contributions
to the extrinsic curvature.
\item[(iv)] The intersection of a four-simplex of type [2,2,1] with $\Sigma_t$
is a tetrahedron. The surface $\hat S$ intersects all four boundary triangles.
However, since there are two types of dual edges which contribute with
angles of opposite sign, the net contribution to the extrinsic curvature vanishes.
\end{itemize}
\begin{table}
        \begin{center}
      \renewcommand{\arraystretch}{1.6}
        \begin{tabular}{ |c|c|c|c|}
                \hline
 simplex  & boundary simplex  &  $\tilde{\rho}$   \\
\hline \hline
$[4,1]$ & $[3,1]$  & $ -{\rm arsinh}\frac{1}{\sqrt{8} \sqrt{1+3 \alpha}} $    \\
                \hline
$[3,2]$ & $[3,1]$  & $  {\rm arsinh}\frac{\sqrt{3}}{2 \sqrt{1+3 \alpha}}  $     \\
                \hline
$[3,2]$ & $[2,2]$  &$  -{\rm arsinh}\frac{1}{\sqrt{6} \sqrt{1+2 \alpha}}$  \\
                \hline
\end{tabular}
\end{center}
\caption{ \small  \label{Winkel3} The angles $\tilde{\rho}$ per simplex type
and per boundary simplex type contributing to the expansion rate 
in 2+2 dimensions. For the time-reversed simplices we have 
$\tilde{\rho}([i,k])=-\tilde{\rho}([k,i]) $.
}
\end{table}

The extrinsic curvature of $\Sigma_t$ with respect to the four-dimensional 
triangulation is concentrated on the two-dimensional faces of the 
$\Sigma_t$-building blocks. 
The angles $\tilde \rho$ needed to calculate this quantity are listed
in Table \ref{Winkel3}. The surface $\hat S$ intersects the two-dimensional 
faces orthogonally, and the extrinsic curvature term $K^{ab}m_{ab}$ appearing 
in the expansion rate (c.f. eq.\ \rf{plm9}) therefore has distributional support on 
these one-dimensional intersections. 
Note that they do not coincide with the dual edges contained in $\hat S$, 
but are positioned transversally to them. 
To determine the integrated extrinsic curvatures $K$ we still need to calculate the lengths 
of these intersections, which is straightforward and will lead to an explicit
$t$-dependence of the expansion rate.
Similarly, to obtain the extrinsic curvatures $k$ one has to multiply the angular 
differences $\Delta\psi$ with the lengths of the associated dual edges.
The final result for the integrated expansion rate for
a $[1_{T,in},1_{T,out},1_{T+1}]$-triangle tower and a surface $\hat S$ in a 
constant-time surface $\Sigma_t$ is given by
\ba \label{221}
\mathfrak{H}_+(\hat S) \!\! &&\!\!\!\!\!\!\! =
a\bigg( 
(1-t)\frac{\pi}{2\sqrt{6}}(N_{131}-N_{311}) +
t \frac{\pi}{3}(N_{122}-N_{212})+ \nn \\
&&\!\!\!\!\!\!\!\!\!\!  \biggl\{ (1-t)(-\sqrt{3})\,{\rm arsinh}(\frac{1}{\sqrt{8}\sqrt{1+3\alpha}})
(N_{311}+N_{131}+\frac{4}{3}N_{221} )
\biggr\} + \nn\\ 
&&\!\!\!\!\!\!\!\!\!\!  3t\, {\rm arsinh}(\frac{1}{\sqrt{6}\sqrt{1+2\alpha}})\, N_{113}+\nn \\
&&\!\!\!\!\!\!\!\!\!\!  \biggl[  ((1-t)  \frac{2}{\sqrt{3}} {\rm arsinh}
(\frac{\sqrt{3}}{2\sqrt{1+3\alpha}}) -2t\,{\rm arsinh}(\frac{1}{\sqrt{6}\sqrt{1+2\alpha}}))
(N_{212}+N_{122})\biggr]\bigg) . \;\,\nn \\
\ea
Similarly, for a $[1_{T-1},1_{T,in},1_{T,out}]$-triangle tower and a 
surface $\hat S$ in a constant-time 
surface $\Sigma_{(T-1)+t}$ one obtains
\ba
\label{222}
\mathfrak{H}_+(\hat S) \!\! &&\!\!\!\!\!\!\! =
a\bigg( 
t\frac{\pi}{2\sqrt{6}}(N_{113}-N_{131}) +
(1-t) \frac{\pi}{3}(N_{212}-N_{221})- \nn \\
&&\!\!\!\!\!\!\!\!\!\!  \biggl\{ t(-\sqrt{3})\,{\rm arsinh}(\frac{1}{\sqrt{8}\sqrt{1+3\alpha}})(N_{131}+N_{113}
+\frac{4}{3}N_{122} )\biggr\} - \nn\\ 
&&\!\!\!\!\!\!\!\!\!\!  3(1-t)\, {\rm arsinh}(\frac{1}{\sqrt{6}\sqrt{1+2\alpha}})\, N_{311}-\nn \\
&&\!\!\!\!\!\!\!\!\!\!  \biggl[ (t  \frac{2}{\sqrt{3}} {\rm arsinh}(\frac{\sqrt{3}}{2\sqrt{1+3\alpha}}) 
-2(1-t)\,{\rm arsinh}(\frac{1}{\sqrt{6}\sqrt{1+2\alpha}}))
(N_{221}+N_{212})\biggr]\bigg)  . \;\, \nn\\
\ea
These formulas have the desired form of depending only on counting-variables
for the simplices in a given triangle tower. As before, one can
achieve a further simplification by adding contributions from successive 
time layers. Consider two triangle towers which are joined by a spacelike 
edge in the base manifold. 
If one chooses the two constant-time surfaces at which the contributions
to the extrinsic curvature are evaluated to be at times $(T\pl t)$ and 
$(T\mi 1)\pl (1\mi t)=(T\mi t)$, the terms in the curly brackets in
\rf{221} and \rf{222} will cancel each other because 
the numbers of the various types of $[4_T,1_{T+1}]$-simplices equal
those of the corresponding $[1_{T-1},4_T]$-simplices.
An obvious and symmetric choice that achieves this cancellation over
a larger number of time steps is therefore to evaluate the expansion
rate always at half-integer times. 
   
To summarize, the expansion rate in 2+2 dimensions is again a sum of terms coming 
from the extrinsic curvature $k$ of $\hat S$ and of the extrinsic curvature $K$ of $\Sigma_t$. 
The former are already determined by the triangulation of the spacelike edge tower, namely, 
the partitioning of the $[4,1]$-simplices into the subtypes $[3,1,1]$, $[1,3,1]$ and $[2,2,1]$, 
and similarly for the [1,4]-simplices.
In analogy with what happened in the $(2\pl 1)$-dimensional case, the contribution from 
the curvature $K$ is determined by the distributions of the simplicial building
blocks of type $[3,2]$ and $[2,3]$. In a $[1_{T,in},1_{T,out},1_{T+1}]$-triangle tower the 
$[1,1,3]$-simplices defocus light rays and in a $[1_{T-1},1_{T,in},1_{T,out}]$-triangle
tower the $[3,1,1]$-simplices focus light rays in the angular directions. This is similar to 
the behaviour of the $[1,1,2]$- and $[2,1,1]$-tetrahedra in 2+1 dimensions. 
Note also that the contributions from the terms in square brackets 
in \rf{221} and \rf{222} are small, because the (absolute value of the)
prefactor is small ($<0.016$ for $\alpha >7/12$,
which is the range usually considered \cite{ajl4d}).
This is due to the fact that the extrinsic curvature terms coming from the 
$[3,1]$- and the $[2,2]$-boundary simplex have opposite signs.

\subsection*{4. Dynamical triangulation of a black hole}

In this section we will explain how to construct a black-hole geometry in
the formulation of causal dynamical triangulations. This is not only a good
exercise in translating metric data (which still depend on a coordinate choice)
into invariant geometric data, but may also be helpful in deciding on 
the boundary conditions for a path integral over black-hole geometries.
It should be kept in mind that unlike in Regge calculus 
\cite{regge} we are working 
with simplices of {\it fixed} squared edge lengths $a^2$ and $-\alpha a^2$ 
for the spacelike and timelike edges respectively. As is well known, these
fixed building blocks are not well suited for approximating arbitrary smooth
geometries pointwise, as may be desirable in a numerical study of 
{\it classical} Einstein gravity. Instead, as already discussed 
in the Introduction, the application we have in mind here is a
nonperturbative quantum superposition of spacetime geometries,
with classical properties emerging only on sufficiently coarse-grained
scales.

This property of dynamical triangulations prevents us from approximating
a given smooth manifold exactly, even in the limit as the lattice spacing
$a\rightarrow 0$. For example, it is not possible to obtain a tesselation of 
three-dimensional flat euclidean space using equilateral tetrahedra.
Nevertheless, one can usually arrange that geometric quantities --
for example, the curvature scalar --
when integrated or averaged over sufficiently large patches match 
certain prescribed values\footnote{This simple strategy does not work for
quantities which are non-negative, for example, the square $R^2$ of
the Riemann curvature scalar. In such cases, one may need to perform
a finite ``renormalization" or use a more sophisticated way of averaging.
Ultimately any such prescription must be motivated by physical considerations.},
and this is what we will employ in the following. 
 
The explicit construction of a triangulated black hole can be simplified
greatly by an appropriate choice of coordinates in the continuum. 
Since causal dynamical triangulations come with a distinguished
notion of proper time, measuring the invariant distance between hypersurfaces 
of constant time, it is natural to start with coordinates in a proper-time gauge,
where the lapse function is equal to one everywhere.
In the following subsections we will describe the relevant black-hole configuration 
(given by the Kottler solution), identify appropriate coordinates for it, 
and construct a triangulation for the $r$-$t$-plane of this geometry.
The latter serves as base space for a full four-dimensional triangulation 
which can be constructed as a triangulation of product type. 
By its very construction our simplicial rendering of the black-hole geometry
is highly nonunique, and should merely be regarded as an illustration
of how it can be achieved.

\subsubsection*{4.1 The Kottler solution}

Since the method of dynamical triangulations requires a positive (bare) cosmological
constant, the relevant spherically symmetric black-hole configuration is not the
Schwarzschild solution, but the so-called Kottler solution which describes a black
hole in a de-Sitter background with a positive cosmological constant $\Lambda$. 
The line element of the Kottler solution in Schwarzschild-like coordinates $(t,r,\theta,\phi)$ 
is given by \cite{kottler}
\ba\label{kott1}
 ds^2 &\!\!\!\!=\!\!\!\!&-\left(1-\frac{2M}{r}-\frac{r^2}{L^2}\right) dt^2+
                   \left(1-\frac{2M}{r}-\frac{r^2}{L^2}\right)^{-1} dr^2+r^2 d\Omega^2  \nn \\
 &\!\!\!\!=:\!\!\!\!& -f(r)  dt^2+f(r)^{-1} dr^2+r^2 d\Omega^2, 
\ea
where $M$ is a mass parameter, $L$ a length parameter defined by 
$L^2=\frac{3}{\Lambda}$, and $d\Omega^2$ the volume element of the unit two-sphere. 
Apart from the physical singularity at $r\equ 0$ there are two coordinate singularities 
at the positive real roots of $f(r)$ (if $M< L/\sqrt{27}$, as we will assume from now on),
corresponding to the black-hole horizon and the cosmological horizon of the de-Sitter 
background. The latter is positioned at $r\equ L$ if the mass $M$ vanishes. 
The maximal analytical continuation of the spacetime geometry (\ref{kott1}) 
includes infinitely many black-hole and cosmological horizons, but for our
present purposes we will
only be interested in a region inside the cosmological horizon which 
contains a single black hole.   

To facilitate the translation of the metric data of the Kottler solution to a dynamical
triangulation, it is convenient to work in a proper-time gauge, that is,
use coordinates in which the lapse function is equal to $1$. 
A particular set of such coordinates are the Painlev\'e-Gullstrand (PG) 
coordinates $(\tau,r,\theta,\phi)$ \cite{PG,mp}, in terms of which the metric
(\ref{kott1}) assumes the form
\ba\label{kott2}
d s^2 =-d\tau^2+\left(d r+\sqrt{1-f(r)} \,d\tau \right)^2 +r^2 d\Omega^2   .
\ea
We observe that the lapse function is equal to $1$, but the shift $N_r$ in the radial
direction is nonvanishing, $N_r= \sqrt{1-f} $. One can also find coordinates 
with the same constant-time surfaces as in (\ref{kott2}) and $N_r\equ 0$, 
but it turns out that for our discussion keeping the nonvanishing shift function 
is more convenient. 
A remarkable property of the PG-coordinates is the fact that the induced 
constant-time surfaces $\Sigma_\tau$ are flat Euclidean three-spaces. 
Our triangulated black hole will have the form of constant-time surfaces 
represented by three-dimensional spatial triangulations which are 
(approximately) flat and connected to each other using the information 
contained in the shift function $N_r$.

\subsubsection*{4.2 Triangulation of the base manifold}

Next, we will triangulate the geometry induced on the hypersurface $\{\theta=${\it const.}, 
$\phi=${\it const.}$\}$, which is parametrized by a coordinate pair $(\tau,r)$. 
The radial coordinate $r$ is a ``proper-length" coordinate in the same sense in 
which $\tau$ is a proper-time coordinate. 
It is therefore natural to identify (a rescaled version of) $\tau$ with the discrete 
time $T$ inherent in a causal dynamical triangulation and (a rescaled version of) 
$r$ with a discrete coordinate $R$
along the one-dimensional spacelike triangulation $\Sigma_T$. 
We can introduce a discrete radial coordinate $R$ simply by taking a
triangulated half-line\footnote{In numerical simulations, there will always
be a maximal radius $R_{max}$, because of the finiteness of the spacetime 
volume.} and enumerating its vertices by 0, 1, 2, ...,
starting from the left-most vertex, say. Since spatial edges by definition
have a geodesic length $a$, discrete and continuous radii are related
by $r=a\, R$. Similarly, one finds $\tau\equ a\sqrt{\alpha\pl1/4}\, T$ for
the relation between the discrete and continuous times.

\begin{figure}[h]
\vspace{0.5cm}
\psfrag{r}{\LARGE{r}}
\psfrag{t}{\LARGE{$\tau$ }}
\centerline{\scalebox{0.52}{\rotatebox{0}{\includegraphics{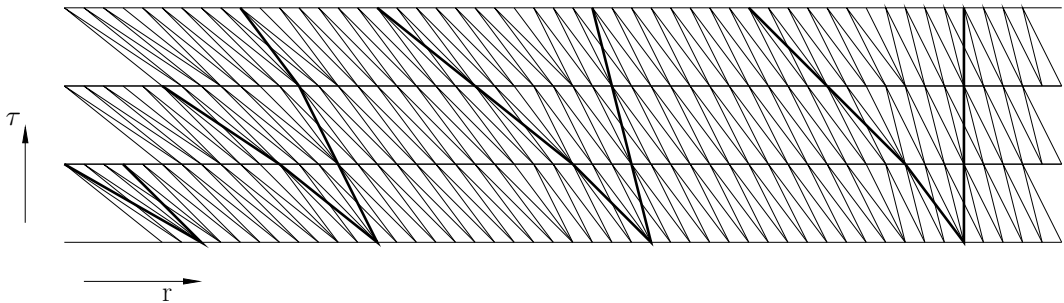}}}}
\caption{ \small \label{schwarzschild} Example of a dynamically triangulated black hole.
The thick lines correspond to light rays, and the vertical light ray on the right
runs along the horizon.}
\vspace{0.5cm}
\end{figure}

Consider now a spacelike edge $e$ in $\Sigma_T$ whose centre has radial 
coordinate $R_T$. If we want to erect a $[2_T,1_{T+1}]$-triangle over this edge 
we must decide which vertex in the next constant-time line $\Sigma_{T+1}$ 
the tip of the triangle should be connected to. 
Because the vector $\vec{\mathbf n}$ pointing from the middle 
of the edge $e$ to the tip of the triangle is normal to the hypersurface 
$\Sigma_T$, the displacement of this top vertex along $\Sigma_{T+1}$ 
from the position $R_T$ on $\Sigma_T$ is determined by the shift $N_r$.
Explicitly, the position $R_{T+1}$ of the top vertex is given by
\be \label{kott3}
R_{T+1}=R_T-N_r(a\,R_T) a^{-1} \Delta \tau\equiv  
R_T-N_r(a\,R_T) \sqrt{\alpha\pl1/4},
\ee
where $\Delta \tau$ is the 
modulus of the length of the vector $\vec{\mathbf n}$. 
In order to construct an explicit triangulation, (\ref{kott3}) 
must be approximated by 
integers. Whichever prescription one chooses, the following cases may occur 
when calculating
the locations of the tips of $[2,1]$-triangles:
\begin{itemize}
\item[(o)]
The normal vectors $\vec{\mathbf n}_1$ and $\vec{\mathbf n}_2$ of two 
neighbouring $[2,1]$-triangles cross within the slice $\Delta T\equ 1$.
For this to happen, $\partial_r N_r$ must be positive and 
$\partial_r N_r >1/\Delta\tau$ i.e. the geometry must vary on scales 
smaller than $a$. For the Kottler solution this does not occur as long as 
$a\ll L=\sqrt{3/\Lambda}$. 
\item[(i)]
The tips of two neighbouring $[2,1]$-triangles coincide. In this case the 
(component in $r$-direction of the) extrinsic curvature is positive since 
the normals converge toward each other.
\item[(ii)]
The tips of two neighbouring $[2,1]$-triangles with base in $\Sigma_T$ are one 
or more edges apart in $\Sigma_{T+1}$. In this case, we fill the space between 
the two $[2_T,1_{T+1}]$-triangles with the appropriate number of 
$[1_T,2_{T+1}]$-triangles. The extrinsic curvature is zero when a
single upside-down
triangle is inserted, and negative otherwise.
\item[(iii)]
The position $R_{T+1}$ assumes negative values. For the Kottler solution 
this happens whenever the discrete radial coordinate $R_T$ satisfies
\be \label{kott4}
(R_{T})^3<\frac{1}{a^3}\ \frac{2M(\Delta\tau)^2}{1-\frac{(\Delta\tau)^2}{L^2}}\approx 
\frac{2 M (\Delta\tau)^2 }{a^3} = \frac{2M}{a} (\alpha +\frac{1}{4}),
\ee
where the approximation in the second step is justified because of $a\ll L$. 
In this case we 
simply omit the $[2,1]$-triangles. This way one obtains a spacetime boundary which
is effectively spacelike (because there are far more spacelike than timelike edges),
as illustrated by Fig.\ \ref{schwarzschild}.
It corresponds to the spacelike singularity of the Kottler solution at $r\equ 0$. 
\end{itemize}
Fig.\ \ref{schwarzschild} shows the typical features of a dynamically 
triangulated black hole: 
approaching smaller radii, the tips of the $[2,1]$-triangles get dragged more 
and more toward the centre of the black hole (left in the figure). 
Similarly, a freely falling particle which starts at some radius $r$ on a surface
$\Sigma_T$ and normal to it, will roughly follow the direction of the 
timelike normals of these triangles and finally fall into the singularity.
The horizon in this picture is located at the radius $r_H$ at which the line 
$r\equ{\it const.}$ is lightlike. 

\subsubsection*{4.3 Triangulation of the four-dimensional manifold}

Finally, we will sketch how to construct a four-dimensional 
dynamically triangulated black-hole geometry by suitably triangulating the
fibres $F$ over one of the triangulated base spaces $B$ of the previous section.
In other words, one has to provide a triangulation of the triangle towers
$\sigma\times F$ for all triangles $\sigma$ of the base space, as described in
Sec.\ 2.3. Since there is no triangulation which is exactly spherically symmetric,
the spherical symmetry of the Kottler solution can be implemented at most in
an averaged sense. 
We will formulate conditions on the numbers of various simplex
types in the fibres which must be satisfied in order that 
the fibre triangulations can be made homogeneous on large scales $l\gg a$. 
We will not spell out how these building blocks should be distributed over
the fibres in a maximally uniform manner.

Since the spatial hypersurfaces $\Sigma_\tau$ of the metric (\ref{kott2}) are flat, 
the radii $r$ provide a direct measure of the areas of the associated two-spheres 
$\{\tau\equ {\it const.}$, $r\equ {\it const.}\}$. This implies that a vertex tower with 
radial coordinate $R$ should contain (approximately) $c_1 R^2$ triangles, 
where $c_1 \approx 4\pi/A_\Delta$ is of order $1$, and
$A_\Delta$ is the area of a spatial triangle in units of $a$. 
This consideration fixes the number of triangles in a vertex tower and 
therefore the numbers of $[3_{R},1_{R+1}]$- and $[1_{R},3_{R+1}]$-tetrahedra
in a spatial edge tower located between the radii $R$ and $R\pl 1$. What remains to 
be specified is the number of $[2_{R},2_{R+1}]$-tetrahedra in the spatial edge tower. 
This can be done by demanding that the (absolute value of the) integrated intrinsic 
curvature of the edge tower be minimal, i.e. as close to zero as possible.

Fixing thus the numbers of simplices appearing in the triangulated
surfaces $\Sigma_T$ of constant time implies that also  
the numbers of $[4_T,1_{T+1}]$- and $[1_T,4_{T+1}]$-simplices 
(in the $[2_T,1_{T+1}]$- and $[1_T,2_{T+1}]$-triangle towers respectively)
of the four-dimensional triangulation are fixed.
Similarly, the number of $[1_{T,in},1_{T,out},3_{T+1}]$-simplices in a 
$[2_T,1_{T+1}]$-triangle tower is already specified by the volume of the 
vertex tower over the tip of the triangle vertex with time coordinate $T\pl 1$. 
All we are left with are the numbers of $[2_{T,in},1_{T,out},2_{T+1}]$- and 
$[1_{T,in},2_{T,out},2_{T+1}]$-simplices or, equivalently, the numbers 
of $[2_{T},2_{T+1}]$-simplices in the timelike edge 
towers $[1_T,1_{T+1}]$. These numbers can also be determined by demanding a 
vanishing integrated four-dimensional intrinsic curvature scalar, 
see \cite{bddiplom} for details.

\subsection*{5. Summary and outlook}

In this paper we have derived an explicit expression for the 
expansion rate of light rays for a simplicial manifold of the kind that occurs
in the sum over geometries in the causal dynamical triangulations approach
to quantum gravity. Its prime intended use is in a horizon finder in the quantum
theory. The vanishing of the {\it integrated} version of the expansion rate
is an indicator for an apparent horizon in situations where the geometry
along the angular directions is homogeneous at sufficiently large scales.
Our non-integrated expression for the expansion can also be used in
more general situations, but a numerical horizon search would be
considerably more involved, because the expansion rate would have
to be monitored not just as a function of radius and time, but also of
the two angular directions. 

As a first step towards understanding the role of black holes in 
nonperturbative quantum gravity we suggest to investigate the situation
where the geometric configurations are approximately spherically
symmetric. In practice, one would first search for the formation of
an apparent horizon as function of the boundary conditions. This
will involve monitoring the integrated expansion rate 
$\mathfrak{H}_+(\hat{S})$ given in eqs.\ \rf{213} and \rf{214}
as a function of the discrete radius and time coordinates of the
triangulated base manifold. One will have to work out how
the local Monte Carlo moves \cite{ajl4d}
change the counting variables appearing
in $\mathfrak{H}_+(\hat{S})$ in a small neighbourhood of triangle
towers. In case one manages to find evidence for an apparent horizon, 
one will try to understand whether also other large-scale properties of
the geometry match those of a (classical) black hole.
Ultimately, one is of course interested in the quantum deviations from
the classical geometry, especially near the horizon and for very small
radii, to obtain further insights into a possible quantum origin of
(horizon) entropy and a quantum resolution of the central singularity.
These are doubtless ambitious goals,  but with some luck and
ingenuity they may just be within reach of our computational means.

\subsection*{Acknowledgments} B.D. thanks the German National Merit Foundation 
for financial support and the Institute for Theoretical Physics of Utrecht University
for hospitality. R.L. acknowledges support by the Netherlands Organisation for
Scientific Research (NWO) under their VICI program.
\vspace{.6cm}

\subsection*{Appendix}

In this appendix, we define the concept of affine coordinates for flat
simplices, and use them to describe hypersurfaces $\Sigma_t$ of constant time,
as well as the surfaces $S$ used in the computation of the extrinsic
curvatures
in Sec.\ 3.3. For illustrative purposes, we also perform a simple model
calculation for $d\equ 2$. 

For the computation of various geometrical quantities like the angles 
\rf{rhoangle} and \rf{psiangle} appearing in the expansion rate \rf{plm13},
it is very convenient to introduce affine or barycentric coordinates 
(see, for instance, \cite{sorkin}). 
Consider a $d$-simplex $\sigma^d$ embedded into $d$-dimensional Minkowski space. 
Let the vectors $\vec{v}_j$, $j\equ 1,\ldots, d\pl 1$ point from the (arbitrarily chosen) origin 
to the $(d\pl 1)$ vertices of the simplex. 
One can then describe an arbitrary point $P$ in $\sigma^d$ as the centre of mass 
of $(d\pl 1)$ appropriately chosen masses $m^j \geq 0$ located at the vertices of the simplex,
\be 
\label{a1}
P=\sum_{j=1}^{d+1} m^j \vec{v}_j,
\ee
where the sum of the coordinates $m^j$ is normalized to
\be
\label{a1norm}
\sum_{j=1}^{d+1} m^j=1.
\ee
An affine vector can be defined by the difference of two points $P_1$ and $P_2$, 
\be 
\label{a2}
\overrightarrow{P_1 P_2}=\sum_{j=1}^{d+1}(m^j_2-m^j_1)\vec{v}_j.
\ee
The sum of the vector components $y^j:=(m^j_2-m^j_1)$ in \rf{a2} vanishes because of the
normalization \rf{a1norm}. We can therefore replace the vectors $\vec{v}_j$ in (\ref{a2}) by
the affine basis vectors
\be 
\label{a3}
{\mathbf a}_j:=\vec{v}_j-\frac{1}{d+1}\sum_{k=1}^{d+1}\vec{v}_k,
\ee
which are overcomplete due to $\sum_j {\mathbf a}_j\equ 0$. 

In order to express the metric in affine coordinates we still need a basis of one-forms dual to 
$\{{\mathbf a}_j\}$. Because of the overcompleteness of the vector basis, the dual basis 
$\{{\mathbf a}^j\}$ is defined by
\ba
\label{a4}
\langle{\mathbf a}_j,{\mathbf a}^k\rangle=\tilde{\delta}_j^k:=\delta^k_j-\frac{1}{d+1},
\ea
where $\tilde \delta^i_j$ is the projector onto the space of affine coordinates, i.e. it projects 
an arbitrary $(d\pl 1)$-tuple of numbers into one whose sum vanishes and it acts as the 
identity on tuples $(y^1,\ldots,y^{d+1})$ which have a vanishing sum $\sum y^j=0$.

The (Minkowski) metric components of a simplex in affine coordinates take the
form \cite{sorkin}
\be
\label{a5}
\tilde \eta_{ij}=-\frac{1}{2} l^2_{km}\tilde\delta^k_i\tilde\delta^m_j,
\ee
where $l_{ij}^2$ is the squared geodesic length of the edge between the $i$-th and $j$-th vertex.
The metric two-form is then given by ${\mathbf\eta}=\tilde \eta_{ij} {\mathbf a}^i{\mathbf a}^j$. 
Since the $\tilde \delta^i_j$'s on the right-hand side of (\ref{a5}) are projections onto the space of 
affine coordinates they can be replaced by ordinary Kronecker symbols $\delta^i_j$ if (\ref{a5}) is 
contracted with coordinate tuples that already fulfill the affine coordinate condition 
(i.e. sum up to zero). One can check eq.\ (\ref{a5}) by contracting it with the simplex edges 
$e_{(ij)}:=\vec{v}_j-\vec{v}_i={\mathbf a}_j-{\mathbf a}_i$ which gives the expected result
\be
\label{a5b}
{\mathbf\eta}(e_{(ij)},e_{(ij)})=l^2_{ij} .
\ee

With the help of the barycentric coordinates one can give an easy characterization of 
the fibres with respect to the two different product structures introduced in Sec.\ 2, namely, the
constant-time hypersurfaces $\Sigma_t$ and the surfaces $S$ of codimension two. 
First, consider a $d$-simplex of type $[N_T,N_{T+1}]$, that is, a simplex having $N_T$ vertices 
in the hypersurface $\Sigma_T$ and $N_{T+1}$ vertices in the neighbouring surface $\Sigma_{T+1}$. 
The barycentric coordinates $(m^1,\ldots,m^{d+1})$ of any point in the intersection of 
$\Sigma_{T+t}$, $0\leq t\leq 1$, and the $d$-simplex $[N_T,N_{T+1}]$ satisfy
\be 
\label{a6}
\sum_{j=1}^{N_T} m^j=1-t=1-\sum_{j=N_T+1}^{d+1} m^j   .
\ee
This condition describes a linear subspace whose points have a ``constant distance" 
from the $T$- and the $(T\pl 1)$- vertices. 

A surface $S$ can be characterized in an analogous manner. Consider a $d$-simplex of type $[N_{T,in},N_{T,out},N_{T+1}]$, with $N_{T,in}$ vertices in the inner and $N_{T,out}$ in the 
outer vertex tower at time $T$ and $N_{T+1}$ vertices in the vertex tower at time $T\pl 1$. 
(We use ``inner" and ``outer" merely as labels to distinguish between the two vertex towers at
equal time.) Then the intersection of the $d$-simplex with the surface $S$ is the set of points having ``constant distance" to these three set of vertices,
\ba 
\label{a7}
\sum_{j=1}^{N_{T,in}} m^j=r, \qquad 
\sum_{j=N_{T,in}+1}^{N_{T,in}+N_{T,out}} m^j=s, \quad 
\sum_{j=N_{T,in}+N_{T,out}+1}^{d+1} m^j=t,  
\ea
where $r+s+t\equ 1$. Obviously these points also fulfil eq.\ (\ref{a6}), which shows that 
$S$ is a submanifold of $\Sigma_{T+t}$. 

To illustrate the construction, consider the simple case of a two-dimensional Lorentzian 
triangulation containing a spacelike hypersurface $\Sigma_{T+t}$. 
We want to calculate the angles $\tilde \rho$ defined in (\ref{rhoangle}) to compute 
the extrinsic curvature of $\Sigma_{T+t}$. Take a $[2_T,1_{T+1}]$-triangle and introduce 
barycentric coordinates $(m^1,m^2,m^3)$. From eq.\ (\ref{a5}), the components of the 
Minkowski metric in affine coordinates are given by 
\begin{equation} 
\label{a8}
\tilde \eta_{ij}= 
 \left\{ \begin{array}{l} 
\;\;\;0 \quad \quad {\rm for} \; i=j ,\\ 
-\frac{a^2}{2}  \quad \;\,{\rm for}\; e_{(ij)}\; {\rm a\; spacelike\; edge}, \\
\;\;\frac{\alpha a^2}{2}\quad \; {\rm for}\; e_{(ij)}\; {\rm a\; timelike\; edge},
\end{array} \right.
\end{equation}
where we have assumed that the metric is contracted with affine coordinate tuples only.
The tangential vector to $\Sigma_{T+t}$ is $s=c_s(1,-1,0)$, with a normalization constant $c_s$. 
Orthogonal to this and future-directed is the normal vector $n=c_n(-\frac{1}{2},-\frac{1}{2},1)$. 
If we take as the timelike boundary simplex the edge $e_{(1,3)}$, the future-directed 
vector parallel to it is ${\mathbf e}_0=c_0(-1,0,1)$. Lastly, orthogonal to ${\mathbf e}_0$ and 
pointing into the $2$-simplex is the vector
\ba
\label{a9}
\tilde{{\mathbf e}}_1=c_1(\frac{-1-2\alpha}{2\alpha},1,\frac{1}{2\alpha})  .
\ea 
Application of the formulas (\ref{rhoangle}) to the (normalized) vectors 
${\mathbf e}_0$, $\tilde{{\mathbf e}}_1$ and $n$ leads to the angle 
\be
\label{a10}
\tilde \rho_{[2,1]}= -{\rm arsinh}(\frac{1}{2\sqrt{\alpha}})  .
\ee
By an analogous calculation for an ``upside-down" triangle $[1_T,2_{T+1}]$ one obtains the 
corresponding angle 
$\tilde \rho_{[1,2]}= -\tilde \rho_{[2,1]}$.

\end{document}